 \documentclass[manuscript]{aastex61}
\usepackage{graphicx}

\begin{document}

%\title{The acceleration of energetic particles at a coronal shock propagating through a streamer-like magnetic field from the flank}
\title{The Acceleration of Energetic Particles at Coronal Shocks and Emergence of a Double Power Law Feature in Particle Energy Spectra}

%%\correspondingauthor{Xiangliang Kong}
%%\email{kongx@sdu.edu.cn}

\author{Xiangliang Kong}
\affil{Shandong Provincial Key Laboratory of Optical Astronomy and Solar-Terrestrial Environment,
and Institute of Space Sciences, Shandong University, Weihai, Shandong 264209, China; kongx@sdu.edu.cn}
\affiliation{Sate Key Laboratory of Space Weather, Chinese Academy of Sciences, Beijing 100190, China}

\author{Fan Guo}
\affiliation{Los Alamos National Laboratory, Los Alamos, NM 87545, USA}
\affiliation{New Mexico Consortium, Los Alamos, NM 87544, USA}

\author{Yao Chen}
\affil{Shandong Provincial Key Laboratory of Optical Astronomy and Solar-Terrestrial Environment,
and Institute of Space Sciences, Shandong University, Weihai, Shandong 264209, China; kongx@sdu.edu.cn}

\author{Joe Giacalone}
\affiliation{Department of Planetary Sciences, University of Arizona, Tucson, AZ 85721, USA}

\begin{abstract}
%In large solar energetic particle (SEP) events, high-energy particles are believed to be accelerated at shocks driven by coronal mass ejections.
We present numerical modelling of particle acceleration at coronal shocks propagating through a streamer-like magnetic field by solving the Parker transport equation with spatial diffusion both along and across the magnetic field.
We show that the location on the shock where the high-energy particle intensity is the largest, depends on the energy of the particles and on time. The acceleration of particles to more than 100 MeV mainly occurs in the shock-streamer interaction region, due to perpendicular shock geometry and the trapping effect of closed magnetic fields. A comparison of the particle spectra to that in a radial magnetic field shows that the intensity at 100 MeV (200 MeV) is enhanced by more than one order (two orders) of magnitude. This indicates that the streamer-like magnetic field can be an important factor in producing large solar energetic particle events.
We also show that the energy spectrum integrated over the simulation domain consists of two different power laws. Further analysis suggests that it may be a mixture of two distinct populations accelerated in the streamer and open field regions, where the  acceleration rate differs substantially.
Our calculations also show that the particle spectra are affected considerably by a number of parameters, such as the streamer tilt angle, particle spatial diffusion coefficient, and shock compression ratio. While the low-energy spectra agree well with standard diffusive shock acceleration theory, the break energy ranges from $\sim$1 MeV to $\sim$90 MeV and the high-energy spectra can extend to $\sim$1 GeV with a slope of $\sim$2-3.
\end{abstract}


\section{Introduction} \label{sec:intro}
Charged particles can be accelerated to energies beyond a few GeV near the Sun during large solar eruptions such as flares and coronal mass ejections (CMEs) \citep[see reviews,][]{reames99,desai16}.
Large solar energetic particle (SEP) events are usually defined as the proton flux in the $>$10 MeV $GOES$ energy channel exceeding 10 particle cm$^{-2}$ s$^{-1}$ sr$^{-1}$ (particle flux unit, pfu) and there have been about 10 events observed per year during solar maximum. Those events are of particular interest to space weather because of their severe radiation threats to human activities in space.
In some extreme SEP events, so-called ground level enhancements (GLEs), the $\sim$GeV ions can produce sufficient secondary particles detectable by ground-based neutron monitors.
There have been only 2 GLE events recorded in solar cycle 24 \citep{gopalswamy13,gopalswamy18}, in comparison to 16 events in solar cycle 23.
%Meanwhile, the number of large SEP events in this solar cycle is significantly less than that of the previous one \citep{mewaldt15,richardson17}.
Although it is known that the solar activity is weak in this solar cycle, the physical reason for the lack of large SEP and GLE events remains unclear \citep[see, e.g.,][]{gopalswamy13,gopalswamy14,giacalone15,mewaldt15,vainio17}.

In large SEP events, high-energy particles are generally believed to be accelerated by shock waves driven by fast CMEs \citep{reames99,desai16}.
Recent observations have shown that CME-driven shocks can form in the low corona and are capable of accelerating particles to high energies \citep[see, e.g.,][]{bemporad10,ma11,gopalswamy12,gopalswamy13b}.
Type II solar radio bursts are produced by electrons accelerated at shocks driven by solar eruptions \citep{nelson85}, and therefore they are commonly used as an indicator for the formation of shocks.
\citet{gopalswamy13b} measured the leading edges or outermost fronts of CMEs at the starting time of associated Type II burts for 32 events in solar cycle 24, and found that the shock can form at heights below 1.5 $R_\odot$.
%as low as 1.2 $R_\odot$ with a mean value of $\sim$1.4 $R_\odot$.
For GLE events in solar cycle 23, \citet{gopalswamy12} showed that the shock formation heights deduced from the onset of Type II bursts are in the narrow range 1.3 to 1.8 $R_\odot$, with a mean value of $\sim$1.5 $R_\odot$, and the CME heights at GLE particle release range from 1.7 to 4 $R_\odot$, with an average of 3.1 $R_\odot$ for well-connected events \citep[see also,][]{reames09a,reames09b}.

The primary acceleration region along the CME shock, i.e., where most of high-energy particles are produced, is a key issue that remains not well known.
It has been shown that in large gradual SEP events the main source of SEPs can be at both the shock nose \citep[e.g.,][]{cane88,reames97,lario17} and the shock flank \citep[e.g.,][]{kahler16,gopalswamy18}.
The magnetic field configuration is of great importance for determining the primary acceleration region because the acceleration rate strongly depends on the angle $\theta_{Bn}$ between the upstream magnetic field and the shock normal.
In the diffusive shock acceleration (DSA) theory, particles are accelerated more rapidly at a nearly perpendicular shock than at a parallel shock \citep{jokipii87,giacalone05a,giacalone05b}.
%Note, however, that self-excited waves produced by streaming energetic particles may further enhance particle acceleration at parallel shocks \citep[e.g.,][]{lee83,lee05,li03}.
It is known that the magnetic field configuration in the corona is very complicated, consisting of large-scale closed field structures such as coronal loops and streamers.
Therefore, when a CME-driven shock forms in the corona, the shock angles along the nonplanar front can vary significantly as it sweeps through the coronal magnetic field \citep[e.g.,][]{kozarev13,schwadron15}, resulting in different particle acceleration rates \citep[e.g.,][]{giacalone05a,giacalone05b,tylka05,tylka06}.
In this study, we will emphasize the importance of coronal magnetic field configuration for particle acceleration at CME-driven shocks.

In the solar corona, streamers are the most prominent quasi-steady structures and CMEs commonly interact with streamers as they expand or move outward \citep[e.g.,][]{chen10,feng11}.
Recent observations have shown that Type II bursts are often associated with
the interaction between shocks and streamers
\citep[e.g.,][]{reiner03,mancuso04,cho08,feng12,feng13,kong12,kong15,chen14}
or the shock propagating through high-density coronal loops
\citep[e.g.,][]{pohjolainen08,cho13,kumar16}.
\citet{kong15,kong16} investigated the effect of a streamer-like magnetic field on electron acceleration at an outward-propagating coronal shock using test-particle simulations, and found that the primary electron acceleration region shifts as the shock propagates and the electron acceleration is more efficient when electrons are trapped in the closed field lines compared to electrons injected in open field regions.

%due to limited capability of current coronagraph observations of CME-streamer interaction in the corona, and the unavailability of direct measurement of coronal magnetic field topology,
However, it remains unclear what is the effect of streamers or CME/shock-streamer interaction on the acceleration, trapping and release of SEPs \citep[see, e.g.,][]{kahler00,kahler05,rouillard16,kocharov17}.
\citet{kocharov17} made detailed multi-wavelength analysis of two GLE events, and suggested that relativistic protons initially accelerated during the flare and CME launch could be trapped in large-scale magnetic loops and later released by the expansion of CME and interaction of the CME with coronal streamers.
A CME may originate below or outside of a streamer depending on the location of its active region, therefore the primary particle acceleration region (with perpendicular shock geometry) can occur both at the CME nose and flanks.
The two GLE events in this solar cycle occurred on 2012 May 17 \citep{gopalswamy13,shen13,rouillard16} and 2017 September 10 \citep{cohen18,gopalswamy18,guo18,luhmann18,zhao18b}, respectively.
In the first event the CME originated below the streamer belt \citep{rouillard16}, while in the second event the CME eruption region is located outside and near the edge of the streamer belt \citep{luhmann18}.
Observations of CME/shock-streamer interaction for the two GLE events are presented in the Appendix.

Recently, \citet{kong17} presented a numerical model to investigate particle acceleration at CME-driven shocks close to the Sun, by considering a coronal shock with a kinematic description propagating through a streamer-like magnetic field that is analytically given \citep{low86}.
Particle acceleration is studied by solving the Parker transport equation \citep{parker65} with both parallel and perpendicular diffusion.
They showed that particles can be sufficiently accelerated to above 100 MeV within 2-3 $R_\odot$, and when the shock propagates through a streamer-like magnetic field, particles are more efficiently accelerated due to perpendicular shock geometry and the natural trapping effect of closed magnetic field lines.
In this study, we further investigate the effect of large-scale streamer-like magnetic configuration on particle acceleration at coronal shocks when the CME-shock originates outside of the streamer and propagates through the streamer from the flank.

The rest of this paper is organized as follows.
%In Section 2, we present observations of CME-streamer interaction for the two GLE events in this solar cycle.
In Section 2, we describe the numerical model.
In Section 3, we present the simulation results including the spatial distribution and energy spectra of accelerated particles. We also examine the effect of various parameters such as the diffusion coefficient on the particle energy spectra via a parameter study.
In Section 4, we discuss the occurrence of double-power-law features in particle energy spectra.
Conclusions are given in Section 5.

%\section{Observations}

\section{Numerical Model} \label{sec:model}

\begin{figure}
\centering
\includegraphics[width=0.6\linewidth]{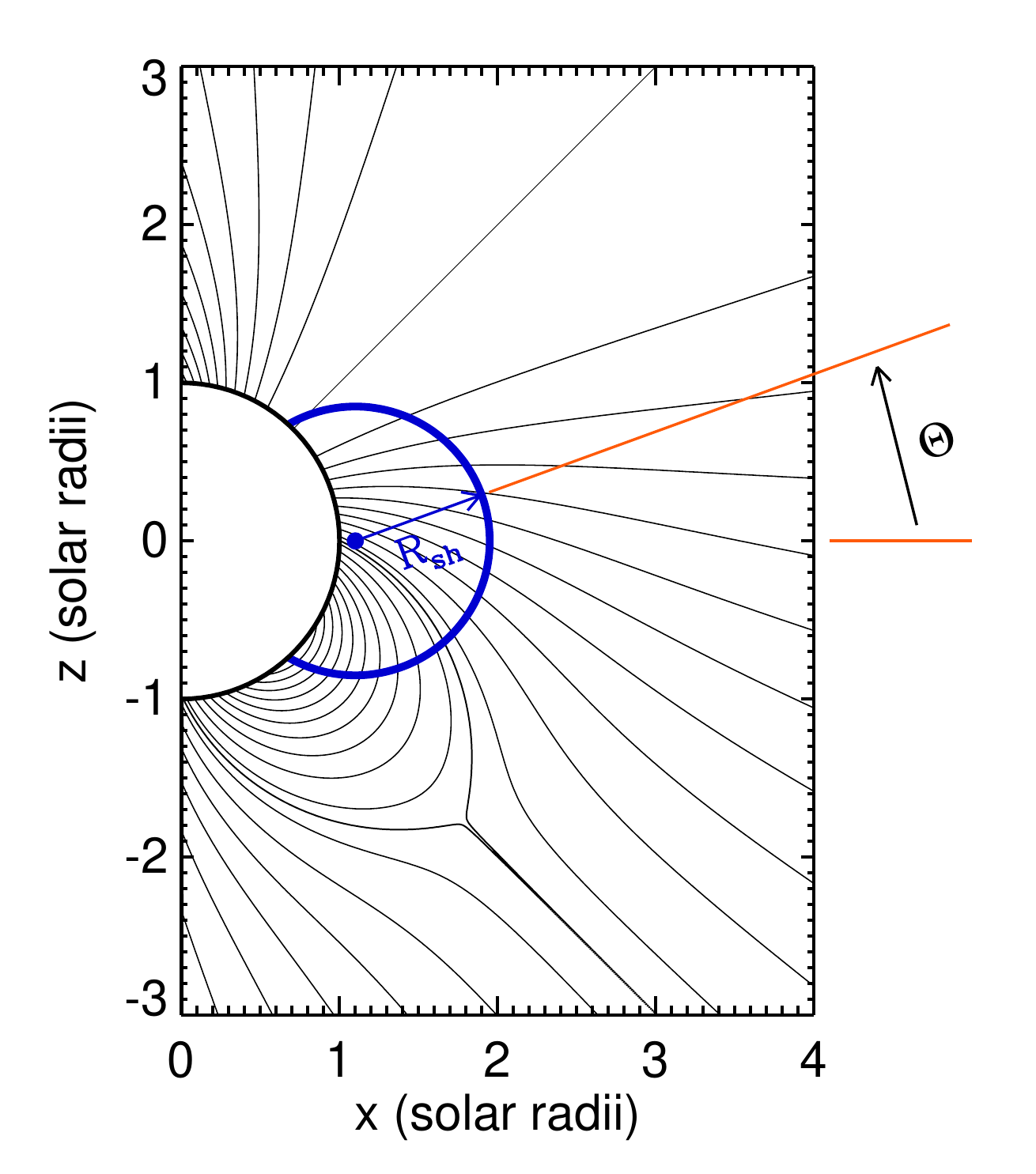}
\caption{
Schematics of the coronal shock morphology (thick blue circle) and the streamer-like coronal magnetic field (black lines). The blue dot indicates the center of the shock, fixed at 0.1 $R_\odot$ above the solar surface.
}
\label{fig:model}
\end{figure}

In this study, we further investigate the effect of large-scale streamer-like magnetic configuration on particle acceleration at coronal shocks using the numerical model developed by \citet{kong17}.
A schematic of the numerical model is shown in Figure \ref{fig:model}.
We consider a CME-driven shock originates outside of the streamer and propagates through the streamer from the flank. The outward-propagating coronal shock is represented by an expanding circular front. The background coronal magnetic field is taken to be an analytical solution of a streamer-like configuration \citep{low86}.
In the reference simulation, the streamer-like magnetic field is rotated by 45$^{\circ}$ toward the solar south pole, referred to as the streamer tilt angle $\Theta_{tilt}$. We also examine the effect of $\Theta_{tilt}$ by considering the cases $\Theta_{tilt}$ = 30$^{\circ}$ and 60$^{\circ}$.
The acceleration of protons in the shock-streamer system is modelled by numerically solving the Parker transport equation \citep{parker65} using a stochastic integration method \citep[e.g.,][]{guo10,senanayake13,li18}.

As shown in Figure \ref{fig:model}, we use two coordinate systems simultaneously. In the helicocentric Cartesian coordinate system ($x$, $y$, $z$),  the origin is at the center of the Sun, with the $z$-axis being the solar rotation axis and the $x$-axis being toward the intersection of the solar equator and solar central meridian.
Note that the analytic solution of the coronal magnetic field is axisymmetric about the solar rotation axis \citep{low86}. We only consider a two-dimensional simulation in the $x$-$z$ plane.
In the shock coordinate system ($R$, $\Theta$) in the $x$-$z$ plane, the origin is set to be at the center of the shock. The $R$ direction points outward radially from the shock center (i.e., along the shock normal), and $\Theta$ is defined as the angle with respect to the $x$ axis and ranges from $-\pi/2$ to $\pi/2$.

We assume a circular shock front forms with an initial radius of 0.2 $R_\odot$ (at $t$ = 0) and moves outward with a constant speed $V_{sh}$ = 2000 km s$^{-1}$. The center of the shock is fixed in the solar equatorial plane at a height of 0.1 $R_\odot$ above the solar surface, which is a reasonable assumption in the low corona region since the expansion is usually dominant in the initial phase of CMEs \citep[e.g.,][]{kwon14,liu19}.
In recent observational studies, the density compression ratios along the shock fronts were deduced to be in the wide range of 1 to 3 \citep[e.g.,][]{bemporad10,bemporad14,susino15,kwon18}.
For the reference run, the shock compression ratio $X$ is 3.
We also examine the effect of varying $X$ and consider two other values, 2 and 2.5, respectively.

We describe the sharp variation in the shock layer by a hyperbolic tangent function. In local shock frame, the fluid velocity along the shock normal is given by,
\begin{equation}
 U(x') = \frac{U_1 + U_2}{2} - \frac{U_1 - U_2}{2} tanh \left(\frac{x'}{\delta_{sh}} \right),
\end{equation}
where $x'$ is the distance to the shock front along the shock normal, $U_1$ and $U_2$ are the upstream and downstream normal flow speeds, and $\delta_{sh}$ is the width of the shock.
To obtain the correct solution of DSA, the width of the shock $\delta_{sh}$ should be much smaller than the characteristic diffusion length at the shock front $L_{d} = \kappa_{nn} /V_{sh}$, where $\kappa_{nn}$ is the diffusion coefficient in the direction normal to the shock. We take $\delta_{sh}$ = 4 $\times$ 10$^{-5}$ $R_\odot$ in our simulations to ensure that this condition is satisfied.
Note that the shock thickness resolved in our approach is much smaller than the cell size of MHD simulations which are used to provide flow and magnetic fields for the calculation of particle acceleration in a similar way \citep[e.g.,][]{kozarev13,schwadron15}.

The magnetic field is compressed when it is swept by the shock into the downstream region. In the shock layer (within 8 $\delta_{sh}$), the magnetic field is determined by the MHD shock jump conditions in the local shock frame. The transverse component of the magnetic field with respect to the shock surface is given by $B_{\Theta} = B_{\Theta 0} V_{sh}/U(x')$, where $B_{\Theta 0}$ is the transverse component of the background field and $x'$ is the distance to the shock front along the shock normal. The normal component of the magnetic field $B_R$ remains unchanged.
In the downstream region, the magnetic field is obtained from the ideal MHD induction equation \citep{kong17}. While our two-dimensional treatment is simplistic and leads to non-zero divergence of magnetic field at the shock front, we argue that particle acceleration and transport mainly depend on large-scale magnetic field geometry. However, we do acknowledge that the errors in magnetic field within the transition layer can mismatch parallel diffusion and perpendicular diffusion, and may influence our results. In the future three-dimensional studies, we will use more sophisticated treatment for ensuring the exact solution of Maxwell's equations \citep[e.g.,][]{giacalone17}.

For diffusion coefficients used in the Parker transport equation, we first obtain particle diffusion both along and across the magnetic field for the reference run from a few considerations, and then vary them and see how the key results change.
The diffusion coefficient tensor is given by,
\begin{equation}
\kappa_{ij} = \kappa_{\perp} \delta_{ij} + (\kappa_{\parallel} - \kappa_{\perp}) \frac{B_i B_j}{B^2},
\end{equation}
where $\kappa_{\parallel}$ and $\kappa_{\perp}$ are the parallel and perpendicular diffusion coefficients, and $B_{i}$ is the average magnetic field vector.
Note that the antisymmetric diffusion coefficient $\kappa_{A}$ related to particle drifts is neglected here because the gradient and curvature drifts are out of the simulation plane.
$\kappa_{\parallel}$ can be calculated from the quasilinear theory  \citep{jokipii71}.
We assume the magnetic turbulence is well developed and has a Kolmogorov power spectrum $P \propto k^{-5/3}$.
When the particle gyroradius is much smaller than the correlation length of turbulence, the resulting diffusion coefficient $\kappa_{\parallel} \propto p^{4/3}$.
The value of $\kappa_{\parallel}$ has the following expression \citep{giacalone99},
\begin{equation}
\kappa_{\parallel} = \frac{3 v^3}{20 L_c \Omega^2 \sigma^2}
csc\left(\frac{3 \pi}{5}\right) \left[1 + \frac{72}{7} \left(\frac{\Omega L_c}{v} \right)^{5/3} \right],
\end{equation}  %}
where $v$ is the particle speed, $L_c$ is the turbulence correlation length, $\sigma^2$ is the normalized wave variance of turbulence, and $\Omega$ is the particle gyrofrequency. The magnetic field and turbulence property in the corona is highly unknown as no in situ observation is available (see \citet{zhao18a} for latest observation of turbulence properties at 1 AU).
For the reference run, we assume the average magnetic field in the simulation domain $B_0$ = 1 G, the turbulence correlation length $L_c$ = 0.01 $R_\odot$, and the turbulence variance $\sigma^2 = \delta B^2/B_0^2$ = 0.14. The assumption of correlation length is similar to models of solar wind heating. For example, \citet{hollweg86} assumed that the correlation length $L_c$ = 7520 (B/1 G)$^{-1/2}$ km ($\sim$0.01 $R_\odot$ for $B_0$ = 1 G) based on the assumption that the correlation length is related to mean flux tube spacing, which in turn is related to the average field strength. Interplanetary observation of magnetic fluctuations suggests that the mean values of turbulence variance can be $\sim$0.4 \citep[e.g.,][]{bavassano82}.
Then, $\kappa_{\parallel 0} = 1.4 \times 10^{17}$ cm$^2$ s$^{-1}$ for the proton initial energy $E_0$ = 100 keV. This value is similar to the choice found in previous works \citep[e.g.,][]{Sokolov2004,kocharov12,giacalone15}.

In the simulations, we normalize the length  $L_0$ = 1 $R_\odot$ = 7 $\times 10^{5}$ km and the velocity $V_0$  = $V_{sh}$ = 2000 km s$^{-1}$. So the normalization of the diffusion coefficient is $\kappa _0$ = $L_0 V_0$ = 1.4 $\times 10^{19}$ cm $^2$ s$^{-1}$, then $\kappa_{\parallel 0}$ = 0.01 $\kappa_0$. We also consider $\kappa_{\parallel 0} / \kappa_0$ = 0.1 ($\sigma^2$ = 0.014) and 0.003 ($\sigma^2$ = 0.4) to examine the effect on particle acceleration.
Test-particle simulations have suggested that the perpendicular diffusion coefficient $\kappa_{\perp}$ is about a few percent (0.02-0.04) of the parallel diffusion coefficient  $\kappa_\parallel$ and their ratio $\kappa_{\perp} / \kappa_\parallel$ is nearly independent of particle energy \citep{giacalone99}.
We take $\kappa_{\perp} / \kappa_\parallel$ = 0.04 in the reference run, and also consider $\kappa_{\perp} / \kappa_\parallel$ = 0.01 and 0.003 to study its effect.

When the shock propagates outward, we inject protons with an initial energy $E_0$ = 100 keV continuously into the immediate region upstream of the shock at a constant rate. In each simulation, a total of 1.2 $\times 10^{5}$ pseudo-particles are injected. We use a particle splitting technique to improve the statistics in the particle distribution at high energies \citep[see, e.g.,][]{giacalone96}. For the reference run, at the end of the simulation, the total number of ``daughter" particles with decreased weight is about six times more than the injected.
The injection energy used here is roughly equal to what is required for the Parker transport equation to describe the acceleration process, i.e., the streaming anisotropy is small enough \citep{giacalone99}.
A particle will be removed from the simulation if it reaches the $z$-axis or hits the solar surface.

As mentioned above, to study the effect of various parameters on particle acceleration, such as the streamer tilt angle ($\Theta_{tilt}$), particle spatial diffusion coefficient (by varying $\kappa_{\parallel 0}/\kappa_0$ and $\kappa_{\perp} / \kappa_\parallel$), and shock compression ratio ($X$), we do a parameter study by including eight simulations in addition to the reference run. A summary of the parameters that are varied in the simulations is listed in Table \ref{tab:parameter}.

\section{Simulation Results} \label{sec:result}

\subsection{Results for Run 1}\label{sec:run1}
\begin{figure}
\centering
\includegraphics[width=0.9\linewidth]{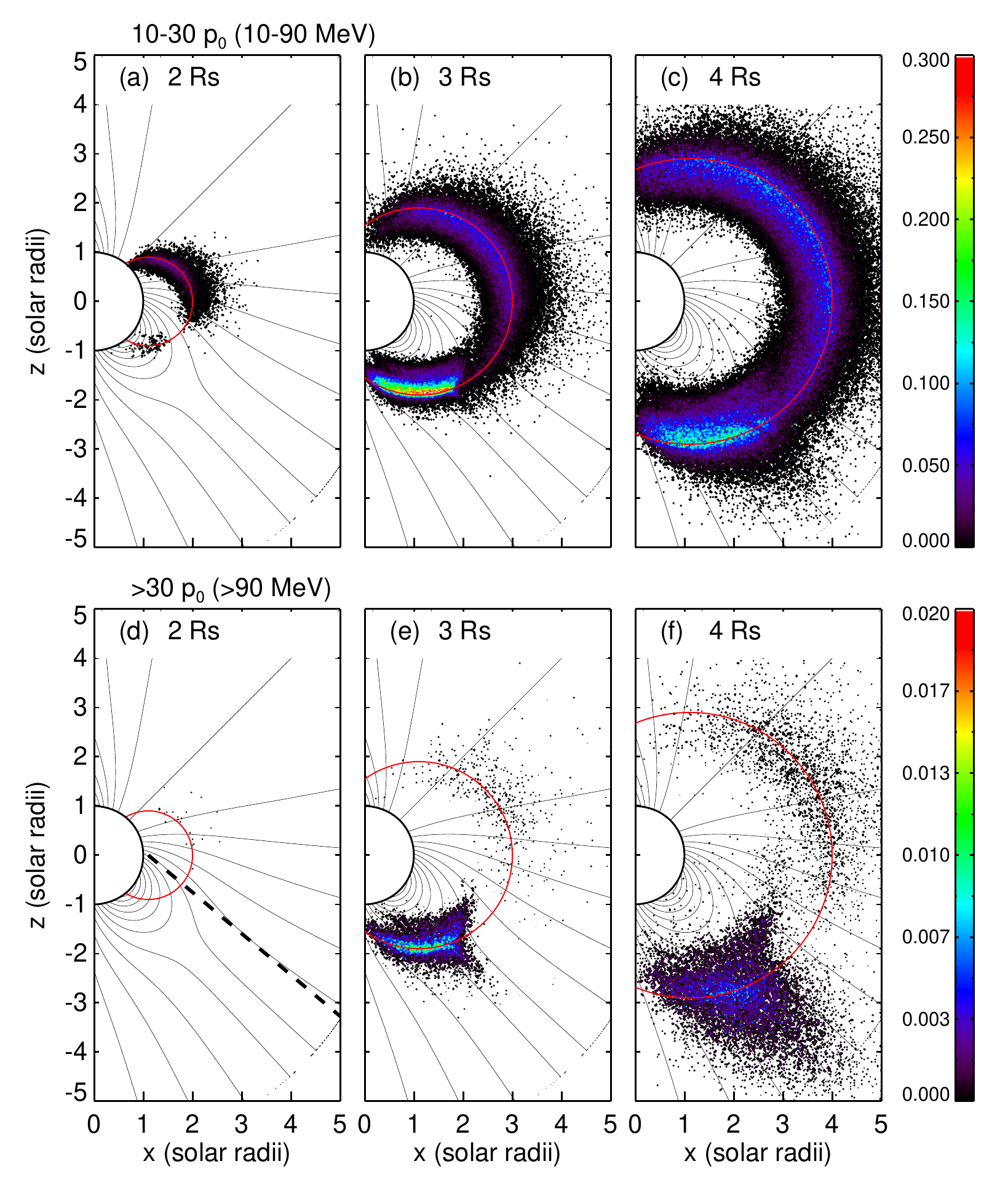}
\caption{
Spatial distributions of accelerated particles with energies 10-90 MeV (upper panels) and $>$90 MeV (lower panels), when the shock reaches three different heights, i.e., 2 $R_\odot$, 3 $R_\odot$, and 4 $R_\odot$, for Run 1. The shock front is denoted by the red circle in each panel. Note that the shock heights in the figures refer to that of the outermost shock front in $x$ direction.
(An animation of this figure is available.)
}
\label{fig:dist-run1}
\end{figure}

We first present the results for the reference simulation, i.e., Run 1 in Table 1.
Figure \ref{fig:dist-run1} shows the spatial distributions of accelerated particles when the shock moves to three different heights, i.e., 2 $R_\odot$ ($t$ = 245 s), 3 $R_\odot$ ($t$ = 595 s), and 4 $R_\odot$ ($t$ = 945 s).
Note that the shock heights in this paper refer to that of the outermost shock front at the equator, which is 1.3 $R_\odot$ in the beginning of the simulation ($t$ = 0).
The particles with relatively lower energies (10-30 $p_0$, 10-90 MeV) are shown in the upper panels and with higher energies ($>$30 $p_0$, $>$90 MeV) are shown in the lower panels.
Note that the associated animation shows the temporal evolution of particle distributions for three energy ranges, including the low-energy range $<$10 MeV not shown in the paper.
The primary sources for particles with different energies are located at different regions and vary significantly as the shock propagates and expands.
As shown in Figure \ref{fig:dist-run1}, and the associated animation, relatively lower-energy (10-90 MeV) particles are initially accelerated at the upper flank of the shock in the open field region where the shock is perpendicular. Very few particles are accelerated to $>$90 MeV when the shock is still below 2 $R_\odot$. Later, energetic particles begin to appear at the lower shock flank in the streamer region. As the shock moves to 3 $R_\odot$, both 10-90 MeV and $>$90 MeV particles have the highest intensities in this region. This indicates that the streamer-like field plays a critical role for particles accelerated to above $\sim$100 MeV, consistent with the results in our previous study \citep{kong17}. While the earlier study \citep{kong17} has shown the shock-streamer interaction region can be a vital acceleration site for high-energy SEPs, the current simulation reveals a more complicated energetic particle distribution when the shock is propagating at a tilted angle compared to the streamer. For rest of the paper we will explore the consequence of this in more details.

\begin{figure}
\centering
\includegraphics[width=\linewidth]{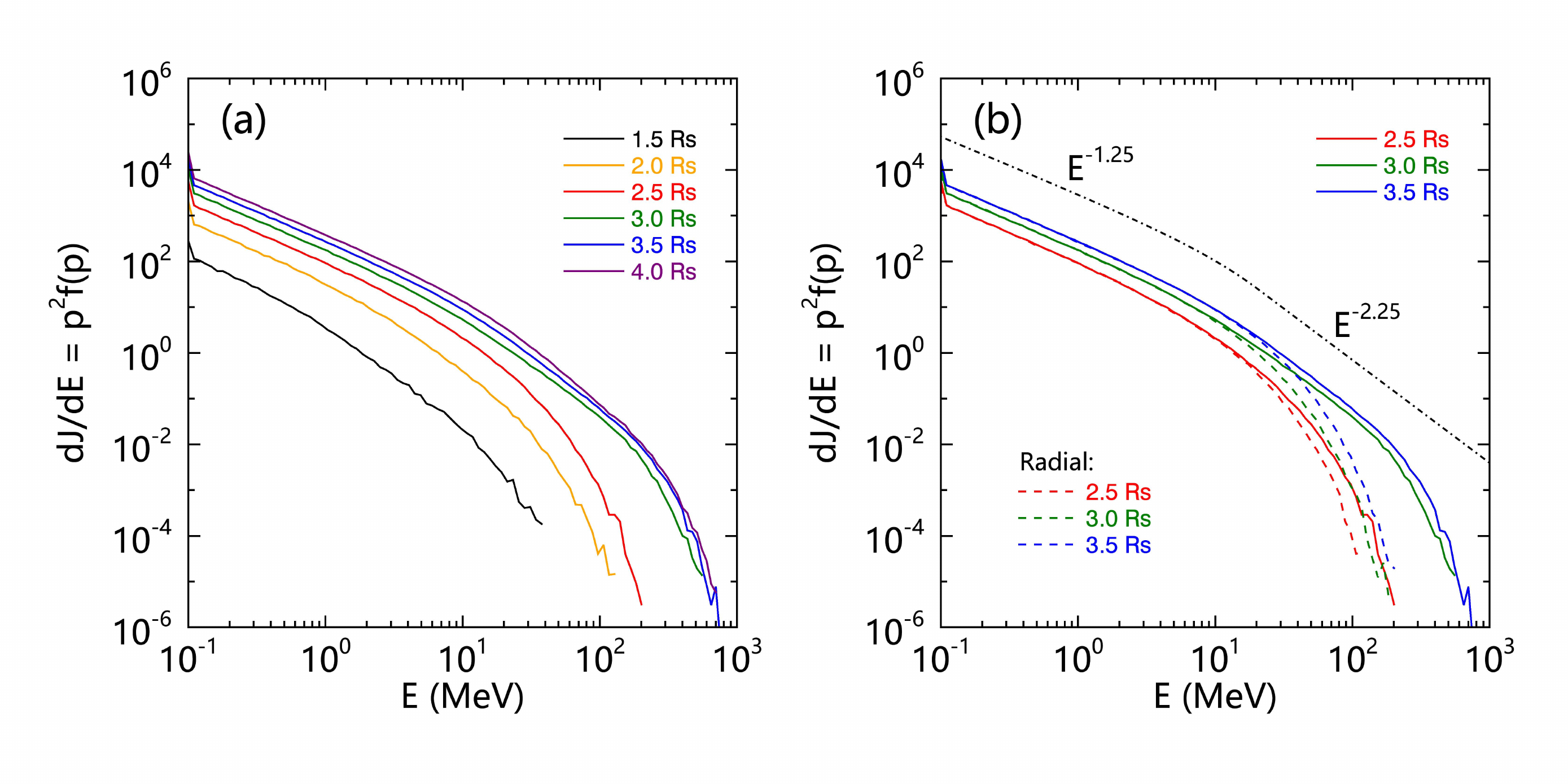}
\caption{
(a) Energy spectra of accelerated particles when the shock propagates to different heights for Run 1.
(b) Comparison of particle spectra of Run 1 with that in a radial magnetic field.
The black dash-dotted line illustrates the fitting of particle spectrum at 3.5 $R_\odot$ (solid blue line) with a double power law function (multiplied by a factor of 10).
}
\label{fig:spec-run1}
\end{figure}

\begin{figure}
\centering
\includegraphics[width=\linewidth]{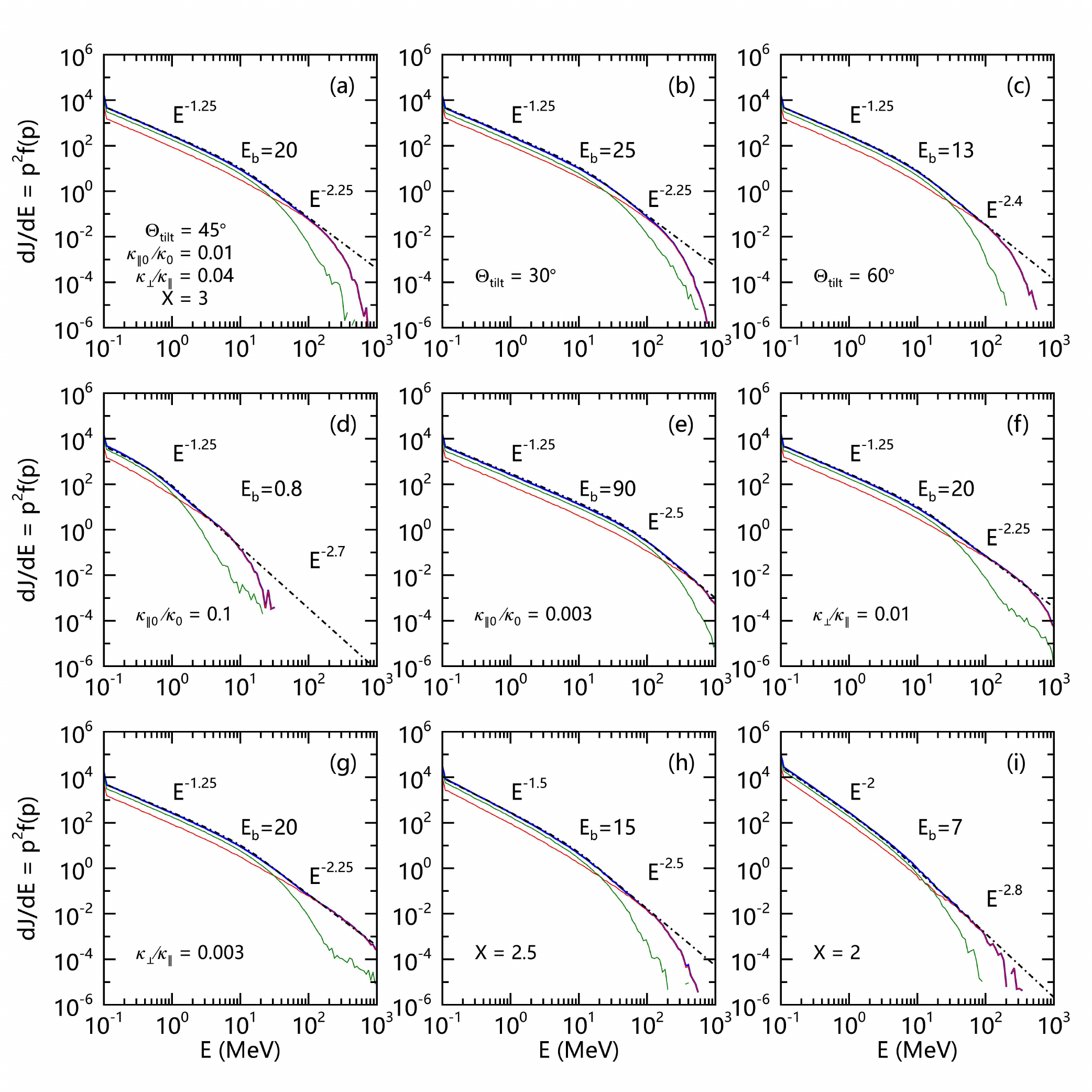}
\caption{
Particle energy spectra when the shock reaches 3.5 $R_\odot$ for all simulations. The parameter that is changed in each simulation compared to that of Run 1 in panel (a) is denoted.
The blue line shows the spectra integrated over the whole simulation domain, while the thin red and green lines show the separated spectra of the streamer region and outside-of-streamer region, as divided by the black dashed line in Figure \ref{fig:dist-run1}.
The integrated spectrum (blue line) in each simulation is fitted with a double power law function, as shown by the black dash-dotted line.
}
\label{fig:spec-9runs}
\end{figure}

For a planar shock with a compression ratio $X$ = 3, the DSA theory predicts a power-law spectrum of particle distribution function $f(p) \propto p^{-3X/(X-1)} = p^{-4.5}$. Therefore, the particle differential intensity $dJ/dE = p^2f(p) \sim E^{-1.25}$ for non-relativistic particles.
Figure \ref{fig:spec-run1}(a) shows the energy spectra of accelerated particles integrated over the whole simulation domain when the shock propagates to six different heights ranging from 1.5 to 4 $R_\odot$.
At low energies, the particle spectra are approximately a power law with a slope of $-$1.25, agreeing well with the slope predicated by DSA theory. At high energies, the spectra break and roll over differently.
Noticeably, the spectrum resembles a double power-law spectrum with a second power-law feature emerges in 10-100 MeV  when the shock front arrives at 3-4 $R_\odot$. We then fit the spectrum according to the Band function form \citep{band93}
\begin{eqnarray}
 dJ/dE & = & C E^{-\gamma_1} exp(-E/E_B)  \quad
            for\ E \leqslant (\gamma_2 - \gamma_1)E_B,  \nonumber \\
       & = & C E^{-\gamma_2} [(\gamma_2 - \gamma_1) E_B]^{\gamma_2 - \gamma_1} exp(\gamma_1 - \gamma_2)           \quad
            for\ E \geqslant (\gamma_2 - \gamma_1)E_B,
\end{eqnarray}
which has four parameters. Here $C$ is the normalization constant, $\gamma_1$ and $\gamma_2$ are the power-law indices for low-energy and high-energy particles, and $E_B$ is the break energy.
The fitted spectrum to the particle spectrum at 3.5 $R_\odot$ (blue line) with the double power law function is shown in Figure \ref{fig:spec-run1}(b) as the dash-dotted line (multiplied by a factor of 10). The fitted parameters are $\gamma_1$ = 1.25, $\gamma_2$ = 2.25, $E_B$ = 20 MeV, and $C$ = 3$\times 10^2$ (see also Figure \ref{fig:spec-9runs}(a)).
Note that in this study we do not try to fit the spectrum in the whole energy range (e.g., the rapid rollover at the high-energy end), but give priority to  a better fit to the low-energy portion.
We will further discuss the emergence of double-power-law feature in particle spectra in Section \ref{sec:dicussion}.

As shown in Figure \ref{fig:spec-run1}(a) and the associated animation, when the shock moves from 2.5 to 3 $R_\odot$, the particle intensity around and above 100 MeV is enhanced dramatically.
During this period, a nearly perpendicular shock geometry forms at the lower shock flank as the shock propagates through the streamer field, therefore resulting in much faster acceleration. In addition, the closed field lines of the streamer also contribute to efficient acceleration by trapping particles upstream of the shock.
The configuration is similar to that occurring at the shock nose as shown in \citet{kong17}.
%\textbf{FG: I commented this section. It seems to be a small detail and I am not sure we want to include it.}
%\textbf{(FG: Looks to me this is just the time for the shock reaching the streamer. Talking about shock height is somewhat misleading because the main acceleration is not happening at the shock nose.) However, as shown in their Figures 2 and 3, particles have been sufficiently accelerated to $>$100 MeV within 2 $R_\odot$.
%This indicates that when the shock originates outside of the streamer, the configuration favorable for rapid acceleration of particles to $>$100 MeV form later, i.e., at higher shock heights, compared to the case when the shock originating below a streamer.}

We repeat the simulation of particle acceleration in a simple radial magnetic field as shown in \citet{kong17}, and compare the particle spectra with Run 1 at three different shock heights in Figure \ref{fig:spec-run1}(b).
The particle spectra for the two simulations are overlapped below 10 MeV, approximately a power law with a slope of $-$1.25.
In the radial magnetic field, the particle spectra at different heights break around 10 MeV and roll over in a similar manner.
In the streamer field case, the maximum energy achievable is much higher, and a significant amount of particles can be accelerated to $>$100 MeV.
At 3-3.5 $R_\odot$, the particle intensity at 100 MeV (200 MeV) is enhanced by more than one order (two orders) of magnitude in Run 1 compared to that in a radial field.
Therefore, particle acceleration at a coronal shock propagating through the streamer-like field is much more efficient than that in a simple radial field. This indicates that the coronal magnetic field configuration can play an important role in producing large SEP events.

\subsection{Effect of the streamer tilt angle}
\begin{figure}
\centering
\includegraphics[width=0.6\linewidth]{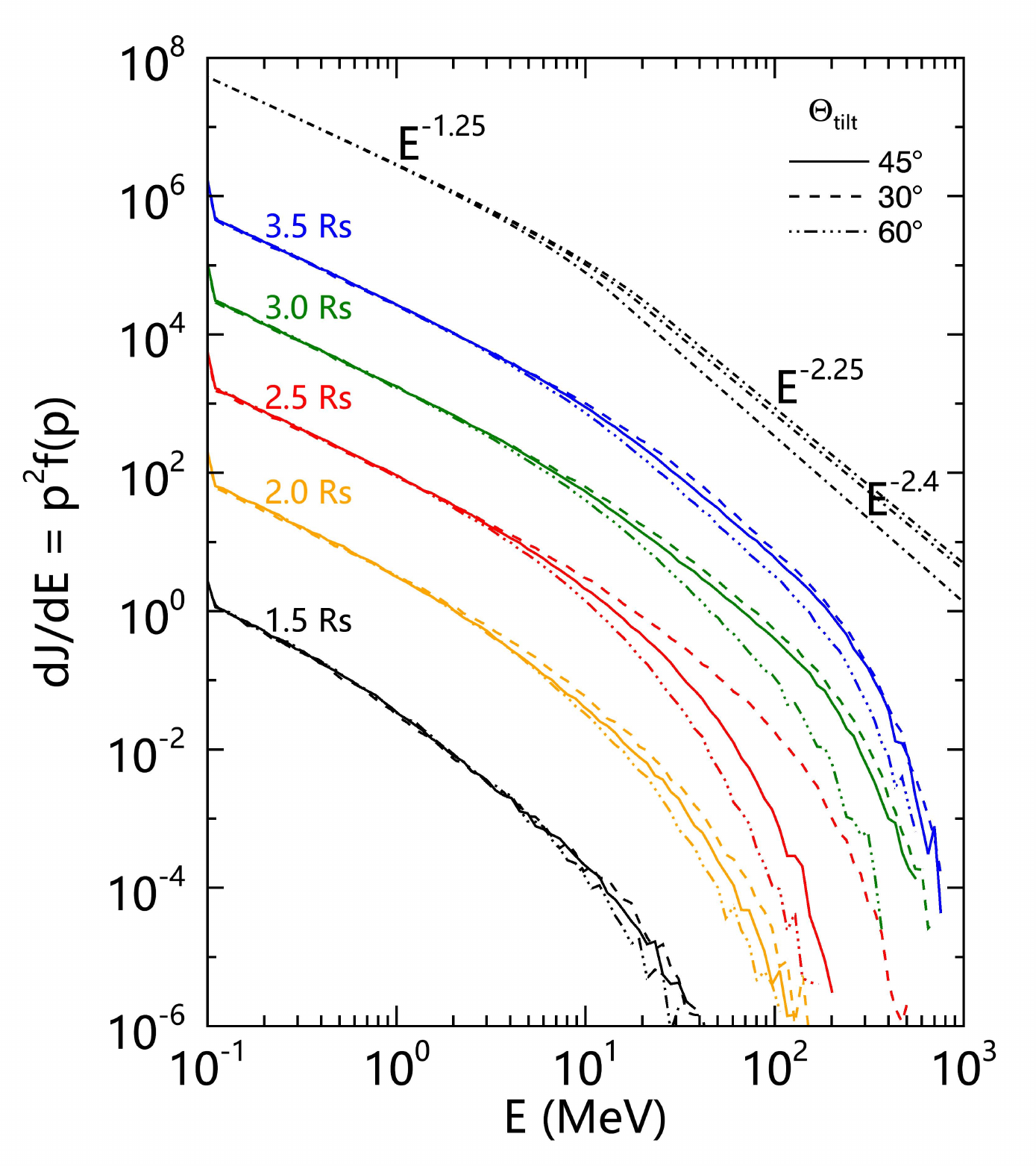}
\caption{
Energy spectra of accelerated particles when the shock propagates to different heights for Runs 1, 2, and 3, with different streamer tilt angles ($\Theta_{tilt}$ = 45$^{\circ}$, 30$^{\circ}$, and 60$^{\circ}$). The spectral profiles at different heights are multiplied by factors of 0.01, 0.1, 1, 10, and 100, respectively.
The black dash-dotted lines indicate the fitting of particle spectra at 3.5 $R_\odot$ (blue lines) with a double power law function (multiplied by a factor of 100).
}
\label{fig:spec-run123}
\end{figure}

A CME-driven shock can interact with the streamer field at different shock heights depending on the shock speed and the distance between the CME eruption region and the streamer.
Thus, the shock geometry and its temporal evolution in the shock-streamer interaction region can vary greatly for different events.
Here we examine the effect of streamer tilt angle ($\Theta_{tilt}$) on particle acceleration by considering other two cases, i.e., the streamer field rotated by 30$^{\circ}$ and 60$^{\circ}$. A comparison of particle spectra for the three simulations, Runs 1, 2, and 3 in Table 1, is shown in Figure \ref{fig:spec-run123}.
At low shock heights, there are no significant difference.
At the shock propagates outward, a rapid hardening in particle spectrum above tens of MeV occurs between 2-2.5 $R_\odot$ in Run 2, between 2.5-3 $R_\odot$ in Run 1, and between 3-3.5 $R_\odot$ in Run 3.
%As discussed above in Section \ref{sec:run1},
This is because the configuration favorable for rapid acceleration of particles start to form later when $\Theta_{tilt}$ is larger.

We fit the particle spectra at 3.5 $R_\odot$ (blue lines) for the three runs with a double power law function, as denoted by black dot-dashed lines (see also Figure \ref{fig:spec-9runs}(a), (b), (c)).
The break energy decreases as increasing $\Theta_{tilt}$ from 30$^{\circ}$ to 60$^{\circ}$, being 25 MeV, 20 MeV, and 13 MeV, for Runs 2, 1, and 3, respectively.
The high-energy power law slope is similar for Runs 1 and 2 ($\gamma_2$ = 2.25), but is slightly steeper for Run 3 ($\gamma_2$ = 2.4).

\subsection{Effect of the value of diffusion coefficient}
The diffusion coefficient describes how well the particles are confined near the shock, therefore has a critical effect on the particle acceleration rate.
The characteristic time for particles to be accelerated from $p_0$ to $p$ at a planar shock is given by \citep[e.g.,][]{drury83},
\begin{equation}
\tau_A = \frac{3}{U_1 -U_2} \int_{p_0}^{p} \frac{dp^\prime}{p^\prime}
\left(\frac{\kappa_{xx,1}(p^\prime)}{U_{x,1}} + \frac{\kappa_{xx,2}(p^\prime)}{U_{x,2}} \right),
\end{equation}
where the subscripts $x$ and $xx$ refer to the direction normal to the shock, 1 and 2 the upstream and downstream regions of the shock.
%$\kappa_{xx}$ is the particle diffusion coefficient in the direction normal to the shock.
Thus, for a smaller diffusion coefficient, the acceleration time is shorter, and the maximum energy attainable will be larger.

\begin{figure}
\centering
\includegraphics[width=0.6\linewidth]{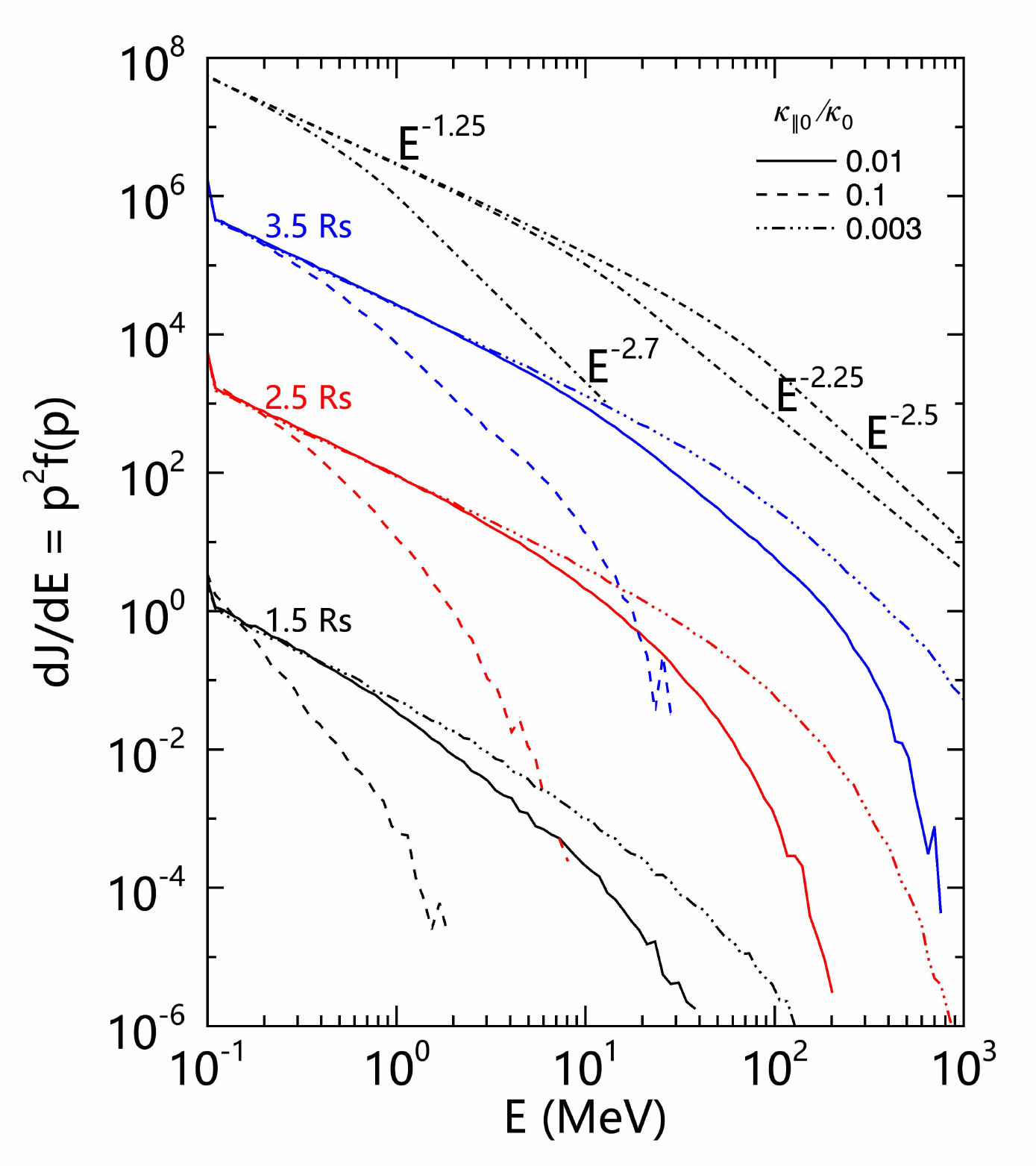}
\caption{
Same as Figure \ref{fig:spec-run123}, but plotted for Runs 1, 4, and 5 with three different values of $\kappa_{\parallel 0}/\kappa_0$ (= 0.01, 0.1, and 0.003).
}
\label{fig:spec-run145}
\end{figure}

We examine the effect of the value of diffusion coefficient by adopting different values of $\kappa_{\parallel 0}/\kappa_0$ and keeping the same ratio $\kappa_{\perp} / \kappa_\parallel$.
Figure \ref{fig:spec-run145} shows the temporal evolution of particle spectra for Runs 1, 4, and 5, with $\kappa_{\parallel 0}/\kappa_0$ being 0.01, 0.1, and 0.003, respectively.
Note that for clarity we only plot particle spectra at three shock heights because the spectral slopes vary greatly for different values of $\kappa_{\parallel 0}/\kappa_0$.
As expected, when the diffusion coefficient is smaller, the particle spectrum is significantly harder.
At 3.5 $R_\odot$, the particle intensity at 100 MeV is enhanced by nearly one order of magnitude for Run 5 compared to Run 1, while the maximum energy in Run 4 is only $\sim$30 MeV.

Same as Figure \ref{fig:spec-run123}, the fittings of particle spectra at 3.5 $R_\odot$ (blue lines) for the three runs with a double power law function are denoted by black dot-dashed lines (see also Figure \ref{fig:spec-9runs}(a), (d), (e)).
The break energy $E_B$ increases when $\kappa_{\parallel 0}/\kappa_0$ is smaller, being 0.8 MeV, 20 MeV, and 90 MeV, for Runs 4, 1, and 5, respectively.
The second power law extends to $\sim$700 MeV in Run 5 ($\kappa_{\parallel 0}/\kappa_0$ = 0.003), compared to only $\sim$10 MeV in Run 4 ($\kappa_{\parallel 0}/\kappa_0$ = 0.1).

%spatial distribution:
%for 0.03, $>$10 MeV particles are mainly accelerated in streamer region, very few particles $>$100 MeV.
%for 0.003, $>$10 MeV particles are roughly uniform, but $>$100 MeV more intense in streamer region.
%the effect of shock geometry decreases with smaller $\kappa$.

\subsection{Effect of the value of perpendicular diffusion}
Perpendicular diffusion plays an important role in particle acceleration at quasi-perpendicular shocks and longitudinal transport of SEPs.
Generally charged particles tend to move following individual field lines.
The diffusion of particles perpendicular to the mean magnetic field includes two forms, the actual crossing of field lines due to scattering or drift and random walking along meandering field lines.
Previous test-particle calculations give a perpendicular diffusion coefficient of about a few percent of the parallel diffusion and nearly independent of particle energy \citep{giacalone99}.
Here we examine the effect of perpendicular diffusion by adopting three different values of $\kappa_{\perp} / \kappa_\parallel$, being 0.04, 0.01, and 0.003, in Runs 1, 6, and 7, respectively.

\begin{figure}
\centering
\includegraphics[width=0.6\linewidth]{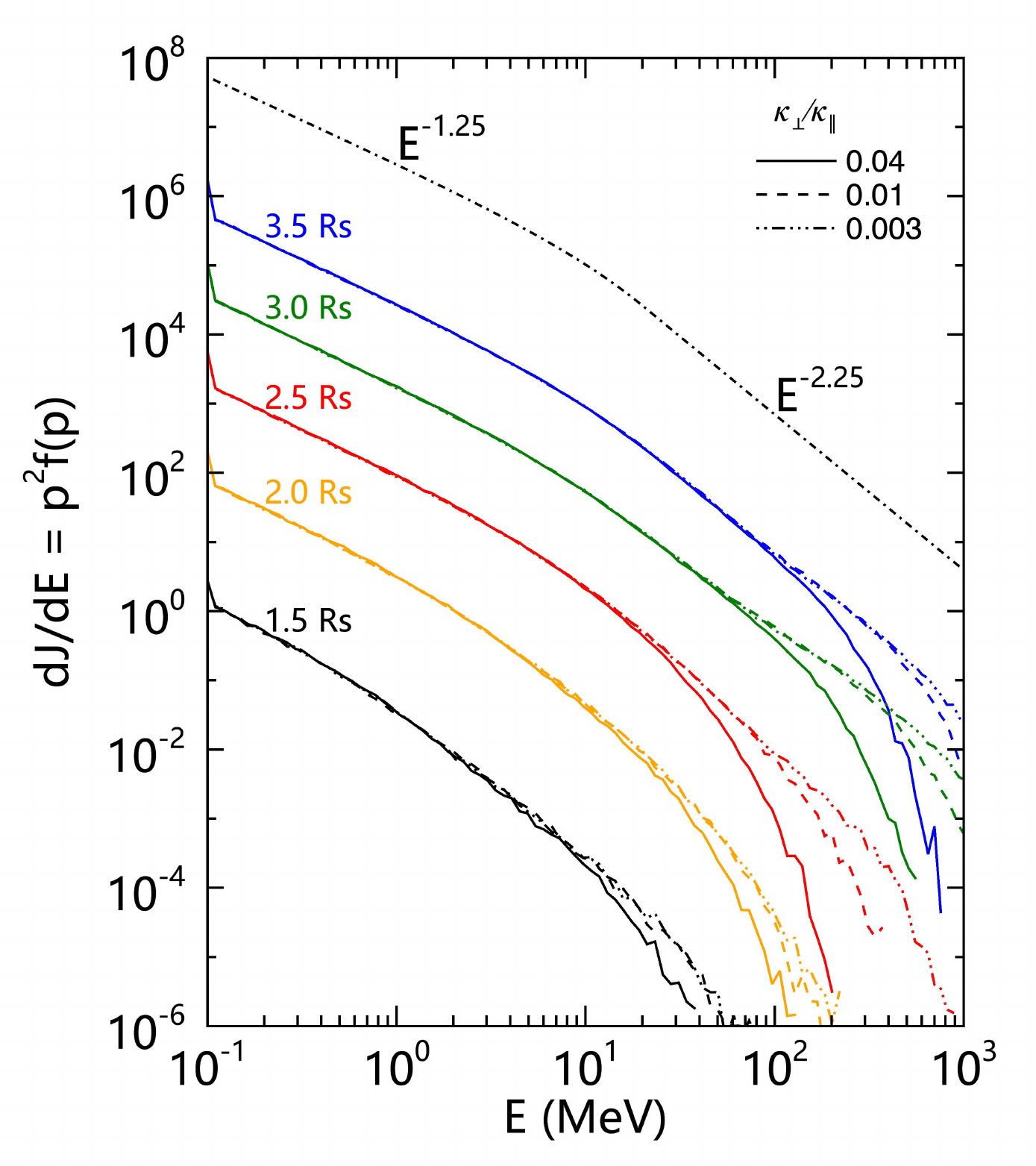}
\caption{
Same as Figure \ref{fig:spec-run123}, but plotted for Runs 1, 6, and 7 with three different values of $\kappa_{\perp} / \kappa_\parallel$ (= 0.04, 0.01, and 0.003).
}
\label{fig:spec-run167}
\end{figure}

Figure \ref{fig:spec-run167} shows the temporal evolution of particle spectra for the three cases.
The high energy portion of the spectra rolls over more slowly in Runs 6 and 7 than in Run 1.
This indicates that the maximum energy of particles is much higher when $\kappa_{\perp} / \kappa_\parallel$ is smaller.
We fit the particle spectra at 3.5 $R_\odot$ (blue lines) for the three runs with a double power law function, as denoted by black dot-dashed lines (see also Figure \ref{fig:spec-9runs}(a), (f), (g)).
The break energy ($E_B$ = 20 MeV) and the high-energy spectral slope ($\gamma_2$ = 2.25) do not vary with $\kappa_{\perp} / \kappa_\parallel$, but the high-energy power law extends to much higher energies, $\sim$400 MeV in Run 6 and $\sim$1 GeV in Run 7, compared to $\sim$100 MeV in Run 1.

\subsection{Effect of the shock compression ratio}
It is known that strong shocks are more rapid accelerators of particles than weak shocks.
In the DSA theory, the power law slope of particle distribution function only depends on the compression ratio ($X$) when applied to a planar shock, i.e., $\alpha = {-3X/(X-1)}$. The particle spectrum is steeper for smaller $X$.
Recent observations have shown that the density compression ratio along the shock front is in the wide range of 1 to 3 \citep[e.g.,][]{bemporad10,bemporad14,susino15}.
We examine the effect of the compression ratio by considering three different values of $X$, being 3, 2.5, and 2 in Runs 1, 8, and 9, respectively.

\begin{figure}
\centering
\includegraphics[width=0.6\linewidth]{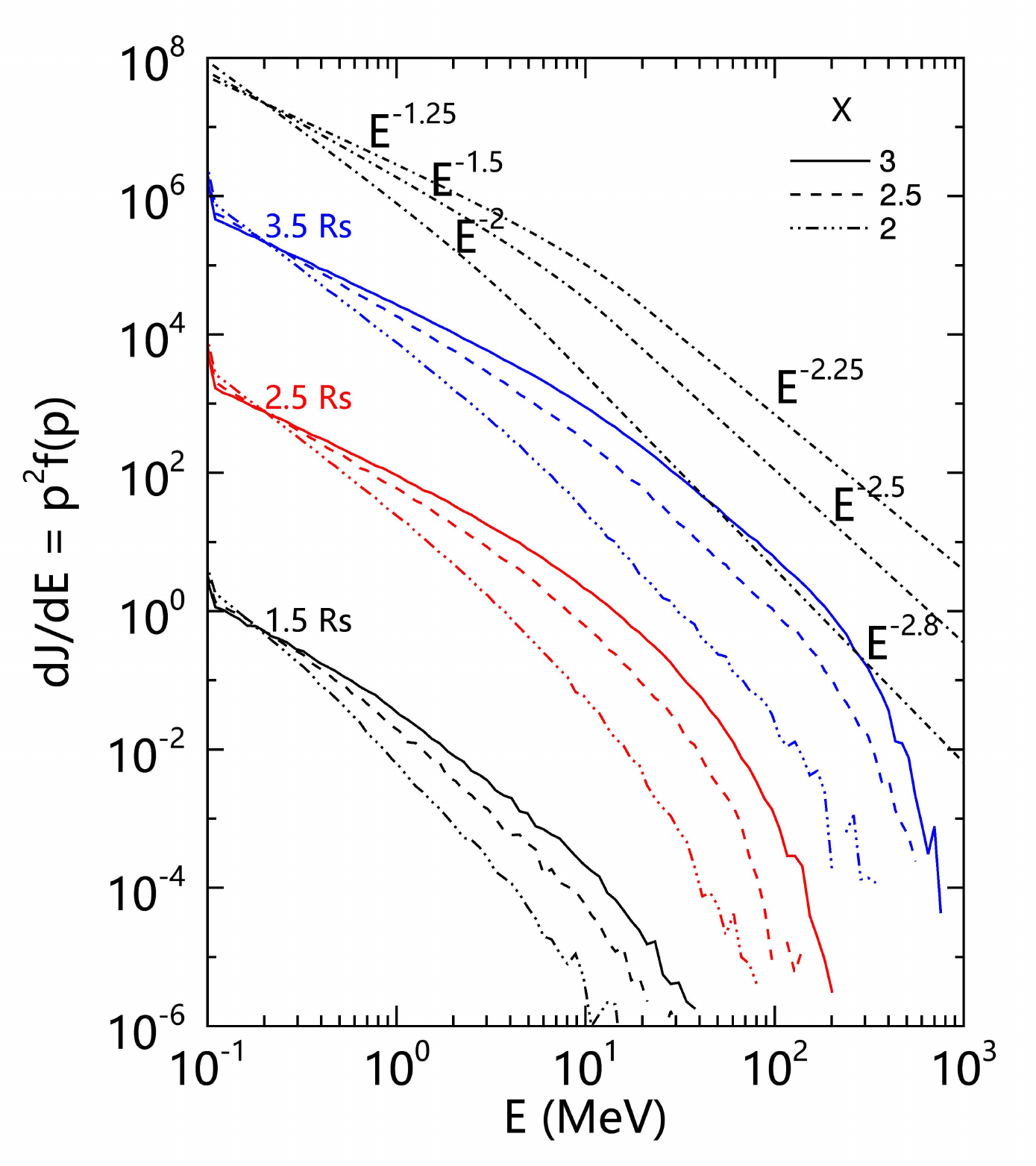}
\caption{
Same as Figure \ref{fig:spec-run123}, but plotted for Runs 1, 8, and 9 with three different shock compression ratios ($X$ = 3, 2.5, 2).
}
\label{fig:spec-run189}
\end{figure}

Figure \ref{fig:spec-run189} shows the temporal evolution of particle spectra for the three cases.
As expected, the particle spectra in Runs 8 and 9 drop more steeply at high energies than in Run 1.
The particle intensity around 100 MeV is declined by nearly one order (two orders) of magnitude in Run 8 (Run 9).
We fit the particle spectra at 3.5 $R_\odot$ (blue lines) for the three runs with a double power law function, as denoted by black dot-dashed lines (see also Figure \ref{fig:spec-9runs}(a), (h), (i)).
The break energy $E_B$ decreases when $X$ is smaller, being 20 MeV, 15 MeV, and 7 MeV, in Runs 1, 8, and 9, respectively. Both the low-energy and high-energy power law slopes get steeper. $\gamma_1$ increases from 1.25 to 1.5 and 2, and $\gamma_2$ increases from 2.25 to 2.5 and 2.8.

\section{Discussion on the Double-power-law Feature in Particle Energy Spectra} \label{sec:dicussion}
%Joe: why there is a second power law, what its slope depends on.
In many large SEP or GLE events, the particle energy spectra exhibit spectral breaks or rollovers at high energies. It was also shown that the break energies depend on the ion charge-to-mass ratio \citep{cohen05,mewaldt05,desai16c}.
The broken spectra are usually described either by a power law with exponential rollover \citep{ellison85} or a double power law \citep{band93}.
\citet{mewaldt12} analyzed 16 GLE events in solar cycle 23 and found that the proton spectra show spectral breaks at energies ranging from $\sim$2 to $\sim$46 MeV and can be well fitted by a double power law up to $\sim$500-700 MeV \citep[see also,][]{desai16b,desai16c,wu18}.
%\citet{desai16b,desai16c} most large SEP events during cycles 23 and 24 are well represented by the double power law.
\citet{cohen18} also found that the particle spectra for proton and other ions in the 2017 September 10 GLE event can be fitted with double power laws and show similar spectral indices (except for proton).
However, \citet{bruno18} showed that the particle spectra between $\sim$80 MeV and a few GeV in 26 major SEP events observed by PAMELA are well described by a power law with exponential rollover, with a spectral slope of 2.2 and break energy at $\sim$170 MeV on average.

%\textbf{
The steady-state DSA predicts a power law spectrum for a planar shock.
However, the physical mechanism responsible for producing spectral breaks, particularly the double power law spectra, is unclear.
%First, at particle acceleration site near the Sun.
Spectral breaks may occur due to effects such as finite acceleration time, finite shock size, shock geometry, and adiabatic cooling \citep[e.g.,][]{zank00,li05,tylka05,tylka06,li09,schwadron15,zhao16a,zhao16}.
\citet{li05} found that a broken power law solution can be achieved if an extra loss term is included in the standard DSA solution.
\citet{schwadron15} obtained broken power-law solutions when considering injection of seed particles in a fixed length along the shock.
%modelled particle acceleration at coronal shocks and showed that broken power-law distributions occur naturally because particles diffuse away and escape from the acceleration sites. \textbf{No. A simple diffusion cannot produce spectral break. It needs something like Gang Li proposed, but it is not clear if that is reasonable.}
%\citet{bruno18} also attributed spectral breaks to particle escaping the shock region during acceleration.
By assuming a single power-law near the Sun, \citet{zhao16} reproduced the double power law spectrum at 1 AU, as a result of pitch-angle diffusion by magnetic turbulence during particle transport in interplanetary space \citep[see also,][]{li15}.

In our numerical model, the acceleration rate varies considerably along the shock front as it sweeping through different magnetic fields.
To understand the origin of double-power-law features in the simulations, we divide the whole simulation domain into two regions, i.e., the streamer region and non-streamer region, as shown by the black dashed line in Figure \ref{fig:dist-run1}(d).
The line is set to be at $\Theta$ = $-40^{\circ}$ ($-30^{\circ}$)
%below the $x$-axis
for runs with streamer tilt angles of 45$^{\circ}$ and 60$^{\circ}$ (30$^{\circ}$).
%\textbf{FG: should we also say it in terms of $\theta$? perhaps label the dashed line using thick and red line?}
It can roughly separate the particles accelerated mainly in the two regions, but at later time particles from the two regions may be mixed together due to cross-field diffusion, as shown in Figure \ref{fig:dist-run1}.

Figure \ref{fig:spec-9runs} shows the particle spectra when the shock moves to 3.5 $R_\odot$ for all simulations. The spectrum integrated for all particles over the whole domain is plotted by the blue line, while the spectra in streamer and non-streamer regions are plotted by thin red and green lines, respectively.
In all cases, the particle spectra from streamer regions have much higher rollover
%\textbf{(FG: roll over usually implies an exponential cut off and break implies a double power-law.)}
energies, but relatively lower intensity in low energy ranges. That is, the low-energy spectra are dominated by particles from non-streamer region, while the high-energy spectra are dominated by particles from streamer regions.
We fit all the integrated spectra (blue lines) with a double power law function, as denoted by the black dotted lines.
It implies that the combination of two particle spectra ordered like these can readily produce a double power law feature in the combined spectra.
In observations, the event-integrated SEP spectra obtained at 1 AU may sample energetic particles accelerated at different locations along the shock due to cross-field diffusion or even from different CME-driven shocks, therefore the double power laws observed in large SEP events may also be affected by the mixing of SEPs from different source regions.

%a combination of different effects

\section{Conclusions} \label{sec:conclusion}
In this study, we have further investigated the effect of large-scale coronal magnetic configuration on particle acceleration at coronal shocks. The shock is represented by an expanding circular front and the coronal magnetic field is given by an analytical solution of a streamer-like configuration \citep{kong17}.
The acceleration of particles at the shock is modelled by numerically solving the Parker transport equation through a stochastic integration method.
We consider a CME-driven shock originates outside of the streamer and propagates through the streamer from the flank. We also conduct a parameter study to examine the effect of various parameters on particle acceleration at coronal shocks, such as the streamer tilt angle, particle spatial diffusion coefficient, and shock compression ratio.

Our calculations show that the primary acceleration regions differ significantly for particles with different energies and vary as the shock propagates. For example, 10-90 MeV particles are initially accelerated at the upper shock flank in the open field and later begin to appear at the lower shock flank in the streamer region, while $>$100 MeV particles are mainly accelerated in the shock-streamer interaction region.
A comparison of the particle spectra to that in a radial magnetic field shows that the particle intensity at 100 MeV (200 MeV) is enhanced by more than one order (two orders) of magnitude in the streamer field.
This indicates that the streamer-like magnetic field can be an important factor in producing large SEP events, consistent with the results in \citet{kong17}. %However, the configuration favorable for rapid acceleration of particles to $>$100 MeV form at higher shock heights, compared to the case when the shock originating below a streamer \citep{kong17}.
%, which affects the particle spatial distribution and energy spectrum
In addition, the particle spectra at 3-4 $R_\odot$ can be well fitted with a double power law up to $\sim$100 MeV. At energies below $\sim$10 MeV, the spectra are a power law with a slope of $-$1.25, agreeing well with the slope predicated by DSA theory. At higher energies up to $\sim$100 MeV, a second power law emerges, with a slope of about $-$2.25. The break energy $E_B$ is around 20 MeV.

Our calculations also show that the particle spectra are affected considerably by a number of parameters.
When the streamer tilt angle $\Theta_{tilt}$ increases from 30$^{\circ}$ to 60$^{\circ}$, the break energy $E_B$ declines from 25 MeV to 13 MeV, and the high energy power law slope is steeper, with $\gamma_2$ varying from 2.25 to 2.4.
When the diffusion coefficient is smaller by reducing $\kappa_{\parallel 0}/\kappa_0$ from 0.1 to 0.003, the particle spectrum gets significantly harder, with $E_B$ increasing from 0.8 MeV to 90 MeV, and the high-energy end of the second power law extending from $\sim$10 MeV to $\sim$700 MeV.
When the perpendicular diffusion is smaller by reducing the ratio $\kappa_{\perp} / \kappa_\parallel$ from 0.04 to 0.003, $E_B$ and $\gamma_2$ do not change, but the high-energy power law can extend to $\sim$1 GeV.
When the shock is weaker by reducing $X$ from 3 to 2, the particle spectra drop more steeply at high energies, with $E_B$ decreasing to 7 MeV, and $\gamma_1$ and $\gamma_2$ increasing to 2 and 2.8, respectively.

The physical mechanisms for double power laws observed in many large SEP or GLE events remain unclear.
Our study shows that a double power law feature emerges readily in particle spectra during acceleration at coronal shocks. Further analysis suggests that it may be a mixture of two distinct populations accelerated mainly in the streamer and open field regions, where the particle acceleration rate differs substantially.
Because the event-integrated spectra observed at 1 AU can sample SEPs from different shock regions or sources due to cross-field diffusion, the mixing effect may also contribute to the formation of double power laws at 1 AU.
It is known that the SEP spectra observed at 1 AU are smeared during interplanetary transport, due to effects such as pitch angle scattering and mixing of different sources.
The recently launched Parker Solar Probe (PSP) will provide observations close to the particle acceleration site near the Sun and help improve our understanding of the acceleration and transport of SEPs and the formation of spectral breaks \citep{desai16}. A clear prediction from the current study is that PSP may observe energetic particles from individual source regions before they are mixed together.

The shock-streamer interaction configuration in this work may also have important implications for sustained $\gamma$-ray emission (SGRE) events, in which $\gamma$-ray emission above 100 MeV can last for hours after the flare impulsive phase
\citep[see, e.g.,][and references therein]{kahler18,klein18,share18}.
The $>$100 MeV $\gamma$-ray emission is thought to be produced by the decay of pions from interactions of $>$300 MeV protons with background ions in the solar atmosphere.
About 30 SGRE events have been observed by $Fermi$ since its launch in 2008, including the two GLE events in this solar cycle.
SGRE events pose a challenge to models of particle acceleration and transport.
Most of latest works favor the CME-driven shock acceleration as the source of high-energy protons, but how the accelerated particles precipitate to the solar photosphere to produce the prolonged $\gamma$-ray emission remain elusive.
As shown in our study, the shock-streamer interaction configuration can provide both efficient acceleration and trapping of energetic particles due to closed field lines of streamers \citep{hudson18,kahler18}.
Future work will focus on the transport of high-energy protons accelerated by the CME-driven shocks and implications for SGRE events.

% If you wish to include an acknowledgments section in your paper,
%% separate it off from the body of the text using the \acknowledgments
%% command.
\acknowledgments
We thank Dr. Mihir Desai for useful comments. %and Dr. Ed Cliver
This work was supported by the National Natural Science Foundation of China (under grants 11873036, 11503014 and 11790304 (11790300)),
the Young Elite Scientists Sponsorship Program by China Association for Science and Technology, the Young Scholars Program of Shandong University, Weihai, and the Project Supported by the Specialized Research Fund for State Key Laboratories. F.G. acknowledges the support from the National Science Foundation under grant 1735414 and support from by the U.S. Department of Energy, Office of Science, Office of Fusion Energy Science, under Award Number DE-SC0018240.
The work of J.G. was supported by NSF under grant 1735422.
The work was carried out at National Supercomputer Center in Tianjin, and the calculations were performed on TianHe-1(A).
We acknowledge the use of GONG magnetic synoptic maps, SOHO LASCO and STEREO SECCHI coronagraph images.

%% Appendix material should be preceded with a single \appendix command.
%% There should be a \section command for each appendix. Mark appendix
%% subsections with the same markup you use in the main body of the paper.
%%\appendix
%%\section{Appendix information}
\appendix
\section{Observations}

\begin{figure}
\centering
\includegraphics[width=\linewidth]{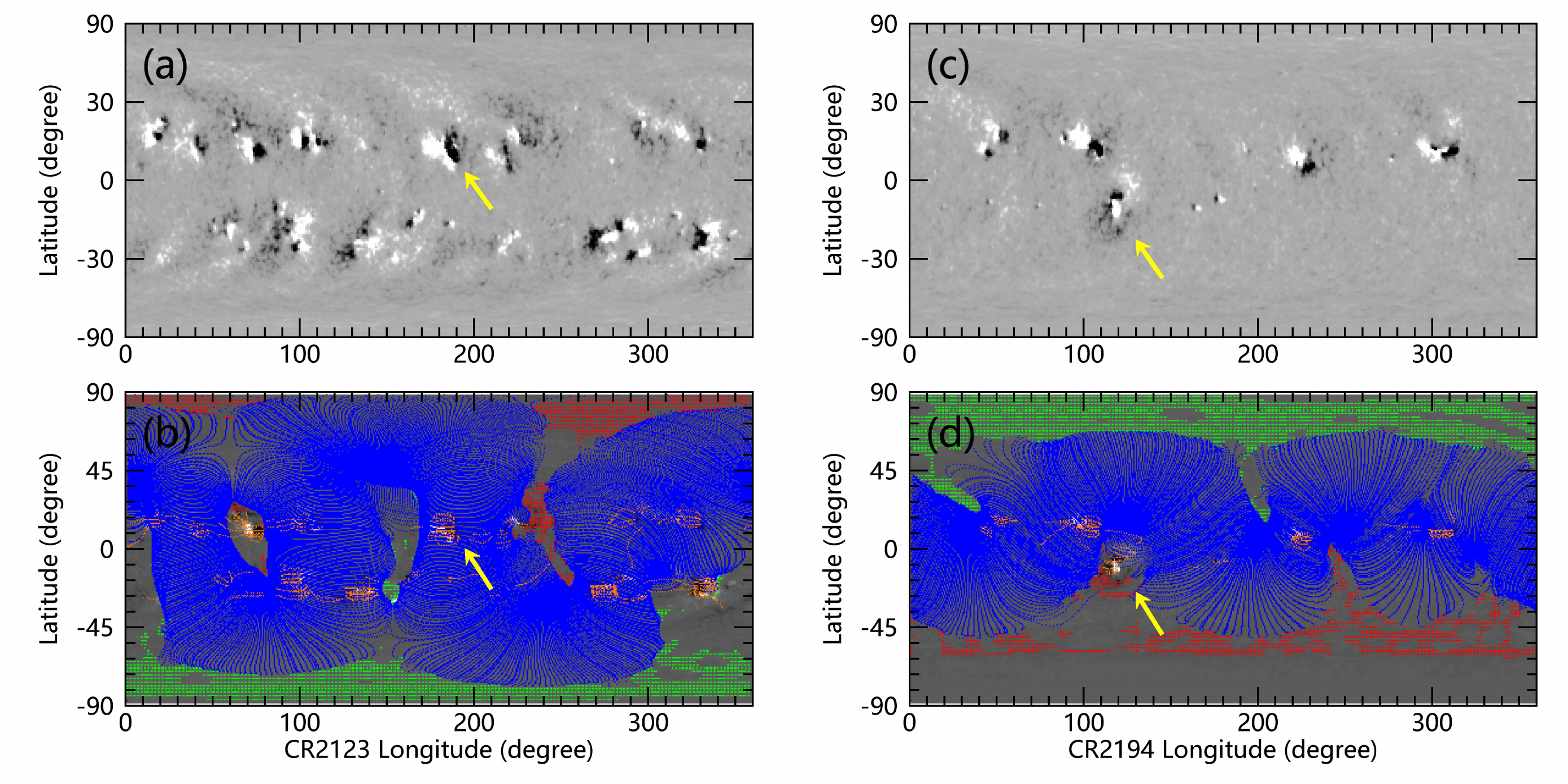}
\caption{
 $GONG$ synoptic magnetic magnetograms (a, c) and coronal magnetic field by PFSS model (b, d) for Carrington Rotations (CRs) 2123 and 2194.
 Four categories of field line are plotted: open positive (outward from the Sun) flux in green, open negative flux in red, the tallest closed flux trajectories (indicating the meandering streamer belt) in blue, and closed active region flux in yellow.
 The yellow arrows point to the active regions of CMEs in the two GLE events on 2012 May 17 (a, b) and 2017 September 10 (c, d), i.e., AR 11476 and AR 12673, respectively.
 }
\label{fig:gongmap}
\end{figure}

The two GLE events in solar cycle 24 occurred on 2012 May 17 and 2017 September 10, respectively.
Figure \ref{fig:gongmap} shows the $GONG$ synoptic magnetic magnetograms and coronal magnetic field by the Potential Field Source Surface (PFSS) model for the two events (\url{https://gong.nso.edu/data/magmap}).
The yellow arrows point to the active regions (ARs) of two CMEs, i.e., AR 11476 and AR 12673.
The streamer belt is illustrated by the tallest closed field lines meandering around the circumference of the Sun in blue in panels (b) and (d).
In the first event the CME originated below the streamer belt \citep{rouillard16}, while in the second event the CME eruption region is located outside and near the edge of the streamer belt \citep{luhmann18}.

\begin{figure}
\centering
\includegraphics[width=0.9\linewidth]{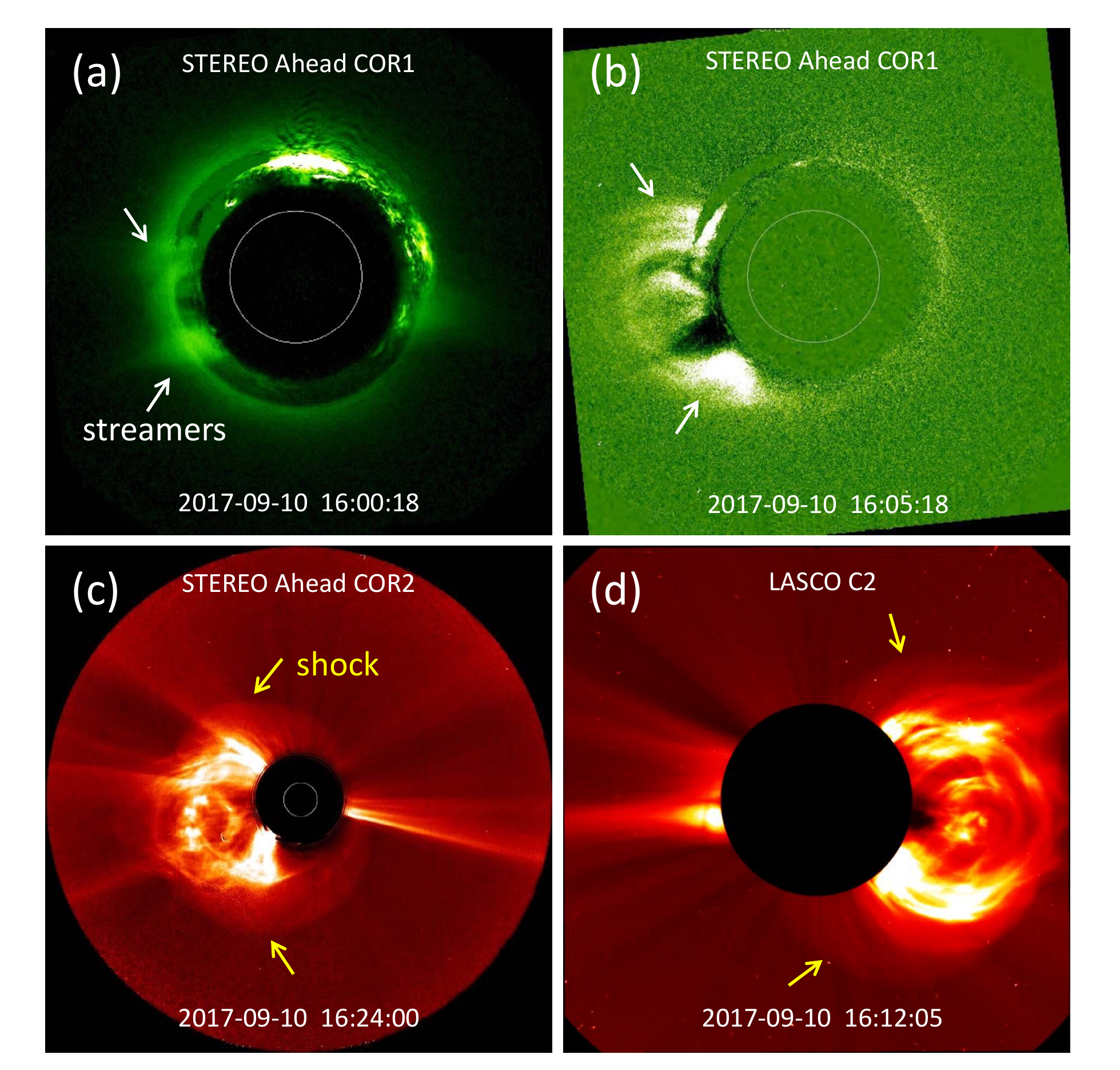}
\caption{
 White-light coronagraph observations of the interaction between CME-shock and streamers for the GLE event on 2017 September 10.
 The white arrows in panels (a) and (b) point to streamers that are strongly deflected by the CME. The yellow arrows in panels (c) and (d) indicate the bubble-shaped structure representing the shock wave front.
}
\label{fig:gle2017}
\end{figure}

Figure \ref{fig:gle2017} shows white-light coronagraph observations of the interaction between CME-shock and streamers for the GLE event on 2017 September 10. The AR12673 is located at the west limb as seen from near-Earth orbit, e.g., in field of view (FOV) SOHO/LASCO.
During this event, the separation angle of STEREO Ahead (STA) with Earth is $\sim$128$^{\circ}$, therefore in STA FOV the AR is located behind the east limb.
In panel (a) at 16:00 UT the CME first appeared in the FOV of STA/COR1, and in panel (b) at 16:05 UT the streamers have been strongly deflected by the CME flanks.
Later, in panels (c) and (d), the outermost CME front shows a bubble-shaped structure in the FOVs of both STA/COR2 and SOHO/LASCO/C2.
As shown in \citet{kwon14}, the outermost faint CME front represents a fast magnetosonic shock wave and can be well reproduced with the ellipsoid model.
The CME-driven shock propagated freely ahead of the CME ejecta, and strongly deflected and passed through the surrounding streamers.

%% The reference list follows the main body and any appendices.
%% Use LaTeX's thebibliography environment to mark up your reference list.
%% Note \begin{thebibliography} is followed by an empty set of
%% curly braces.  If you forget this, LaTeX will generate the error
%% "Perhaps a missing \item?".
%% Note that the style of the \bibitem labels (in []) is slightly
%% different from previous examples.  The natbib system solves a host
%% of citation expression problems, but it is necessary to clearly
%% delimit the year from the author name used in the citation.
%% See the natbib documentation for more details and options.

%%%%%%%%%%%%%%%%%%%%%%%%%%%%%%%%%%%%%%%%%%%%%%%%%%%%%%%%%%%%%%%%%%%%%%%%%

\begin{deluxetable*}{ccccc}[b!]
\tablecaption{Parameters for all simulations \label{tab:parameter}}
\tablecolumns{5}
\tablenum{1}
\tablewidth{0pt}
\tablehead{
\colhead{Run} & \colhead{$\Theta_{tilt}$} & \colhead{$\kappa_{\parallel 0}/\kappa_0$} &
\colhead{$\kappa_{\perp} / \kappa_\parallel$} & \colhead{$X$} \\
%\colhead{}  & \colhead{Tilt Angle} & \colhead{}   %& \colhead{Magnetic Field} & \colhead{Configuration}
}
\startdata
1 & 45$^{\circ}$  & 0.01  & 0.04  &  3 \\
2 & 30$^{\circ}$  & 0.01  & 0.04  &  3 \\
3 & 60$^{\circ}$  & 0.01  & 0.04  &  3 \\
4 & 45$^{\circ}$  & 0.1  & 0.04  &  3 \\
5 & 45$^{\circ}$  & 0.003 & 0.04  &  3 \\
6 & 45$^{\circ}$  & 0.01  & 0.01  &  3 \\
7 & 45$^{\circ}$  & 0.01  & 0.003 &  3 \\
%8 & 30$^{\circ}$  & 0.01  & 0.01  &  3 &\\
%9 & 60$^{\circ}$  & 0.01  & 0.01  &  3 &\\
8 & 45$^{\circ}$  & 0.01  & 0.04  &  2.5 \\
9 & 45$^{\circ}$  & 0.01  & 0.04  &  2 \\
\enddata
\tablecomments{In all simulations, the shock speed $V_{sh}$ = 2000 km s$^{-1}$, %the density compression ratio $X$ = 3,
and the diffusion coefficient has a momentum dependence, $\kappa_\parallel = \kappa_{\parallel 0} (p/p_0)^{4/3}$.}
\end{deluxetable*}

%%%%%%%%%%%%%%%%%%%%%%%%%%%%%%%%%%%%%%%%%%%%%%%%%%%%%%%%%%%%%%%%%%%%%%%%%


\begin{thebibliography}{}
\bibitem[Band et al.(1993)]{band93}Band, D., Matteson, J., Ford, L., et al. 1993, ApJ, 413, 281
\bibitem[Bavassano et al.(1982)]{bavassano82}Bavassano, B., Dobrowolny, M., Fanfoni, G., et al. 1982, SoPh, 78, 373
\bibitem[Bemporad \& Mancuso(2010)]{bemporad10}Bemporad, A., \& Mancuso, S. 2010, ApJ, 720, 130
\bibitem[Bemporad et al.(2014)]{bemporad14}Bemporad, A., Susino, R., Lapenta, G. 2014, ApJ, 784, 102
\bibitem[Bruno et al.(2018)]{bruno18}Bruno, A., Bazilevskaya G. A., Boezio, M., et al. 2018, \apj, 862, 97
\bibitem[Cane et al.(1988)]{cane88}Cane, H. V., Reames, D. V., \& von Rosenvinge, T. T. 1988, JGR, 93, 9555
\bibitem[Chen et al.(2014)]{chen14}Chen, Y., Du, G. H., Feng, L., et al.\ 2014, \apj, 787, 59
\bibitem[Chen et al.(2010)]{chen10}Chen, Y., Song, H. Q., Li, B., et al. 2010, ApJ, 714, 644
\bibitem[Cho et al.(2008)]{cho08}Cho, K. S., Bong, S. C., Kim, Y. H., et al.\ 2008, \aap, 491, 873
\bibitem[Cho et al.(2013)]{cho13}Cho, K.-S., Gopalswamy, N., Kwon, R.-Y., et al.\ 2013, \apj, 765, 148
\bibitem[Cohen \& Mewaldt(2018)]{cohen18}Cohen, C. M. S., \& Mewaldt, R. A. 2018, Space Weather, 16
\bibitem[Cohen et al.(2005)]{cohen05}Cohen, C. M. S., Stone, E., Mewaldt, R. A., et al. 2005, JGR, 110, A09
\bibitem[Desai \& Giacalone(2016)]{desai16}Desai, M. I., \& Giacalone, J.\ 2016, LRSP, 13, 3
\bibitem[Desai et al.(2016a)]{desai16b}Desai, M. I., Mason, G. M., Dayeh, M. A., et al. 2016, ApJa, 816, 68
\bibitem[Desai et al.(2016b)]{desai16c}Desai, M. I., Mason, G. M., Dayeh, M. A., et al. 2016, ApJb, 828, 106
\bibitem[Drury (1983)]{drury83}Drury, L. O. 1983, Rep. Prog. Phys. 46, 973
\bibitem[Ellison \& Ramaty(1985)]{ellison85}Ellison, D. C., \& Ramaty, R. 1985, ApJ, 298, 400
\bibitem[Feng et al.(2012)]{feng12}Feng, S. W., Chen, Y., Kong, X. L., et al.\ 2012, \apj, 753, 21
\bibitem[Feng et al.(2013)]{feng13}Feng, S. W., Chen, Y., Kong, X. L., et al.\ 2013, \apj, 767, 29
\bibitem[Feng et al.(2011)]{feng11}Feng, S. W., Chen, Y., Li, B., et al. 2011, SoPh, 272, 119
\bibitem[Giacalone(2005a)]{giacalone05a}Giacalone, J. 2005a, ApJ, 624, 765
\bibitem[Giacalone(2005b)]{giacalone05b}Giacalone, J. 2005b, ApJ, 628, L37
\bibitem[Giacalone(2015)]{giacalone15}Giacalone, J.\ 2015, \apj, 799, 80
\bibitem[Giacalone(2017)]{giacalone17}Giacalone, J.\ 2017, \apj, 848, 123
\bibitem[Giacalone \& Jokipii(1996)]{giacalone96}Giacalone, J., \& Jokipii, J. R. 1996, JGR, 101, 11095
\bibitem[Giacalone \& Jokipii(1999)]{giacalone99}Giacalone, J., \& Jokipii, J. R.\ 1999, \apj, 520, 204
\bibitem[Gopalswamy et al.(2013a)]{gopalswamy13}Gopalswamy, N., Xie, H., Akiyama, S., et al.\ 2013a, \apjl, 765, L30
\bibitem[Gopalswamy et al.(2014)]{gopalswamy14}Gopalswamy, N., Xie, H., Akiyama, S., et al.\ 2014, EP\&S, 66, 104
\bibitem[Gopalswamy et al.(2013b)]{gopalswamy13b}Gopalswamy, N., Xie, H., Makela, P., et al. 2013b, AdSpR, 51, 1981
\bibitem[Gopalswamy et al.(2012)]{gopalswamy12}Gopalswamy, N., Xie, H., Yashiro, S., et al.\ 2012, \ssr, 171, 23
\bibitem[Gopalswamy et al.(2018)]{gopalswamy18}Gopalswamy, N., Yashiro, S., Makela, P. et al.\ 2018, \apjl, 863, L39
\bibitem[Guo et al.(2018)]{guo18}Guo, J., Dumbovic, M., Wimmer-Schweingruber, R. F., et al. 2018, SpWea, 16, 1156
\bibitem[Guo et al.(2010)]{guo10}Guo, F., Jokipii, J. R., \& Kota, J.\ 2010, \apj, 725, 128
\bibitem[Guo \& Giacalone(2013)]{guo2013} Guo, F., \& Giacalone, J.\ 2013, \apj, 773, 158
\bibitem[Hollweg(1986)]{hollweg86}Hollweg, J. V. 1986, JGR, 91, 4111
\bibitem[Hudson(2018)]{hudson18}Hudson, H. S. 2018, in IAU Symp. 335, Space Weather of the Heliosphere: Processes and Forecasts, ed. C. Foullon \& O. Malandraki (Cambridge: Cambridge Univ. Press), 49
\bibitem[Jokipii(1971)]{jokipii71}Jokipii, J. R. 1971, Rev. Geophys., 9, 27
\bibitem[Jokipii(1987)]{jokipii87}Jokipii, J. R. 1987, \apj, 313, 842
\bibitem[Kahler(2005)]{kahler05}Kahler, S. W. 2005, JGR, 110, A12
\bibitem[Kahler(2016)]{kahler16}Kahler, S. W. 2016, ApJ, 819, 105
\bibitem[Kahler et al.(2018)]{kahler18}Kahler, S.~W., Cliver, E.~W., Kazachenko, M. 2018, ApJ, 868, 81
\bibitem[Kahler et al.(2000)]{kahler00}Kahler, S. W., Reames, D. V., \& Burkepile, J. T. 2000, in ASP Conf. Ser. 206, High Energy Solar Physics Workshop, ed. R. Ramaty \& N. Mandzhavidze (San Francisco, CA: ASP), 468
\bibitem[Klein et al.(2018)]{klein18}Klein, K.-L., Tziotziou, K., Zucca, P., et al. 2018, ASSL, 444, 133
\bibitem[Kozarev et al.(2013)]{kozarev13}Kozarev, K., Evans, R. M., Schwadron, N. A., et al. 2013, \apj, 778, 43
\bibitem[Kocharov et al.(2017)]{kocharov17}Kocharov, L., Pohjolainen, S., Mishev, A., et al. 2017, ApJ, 839, 79
\bibitem[Kocharov et al.(2012)]{kocharov12}Kocharov, L., Vainio, R., Pomoell, J., et al. 2012, ApJ, 753, 87
\bibitem[Kong et al.(2015)]{kong15}Kong, X. L., Chen, Y., Guo, F., et al.\ 2015, \apj, 798, 81
\bibitem[Kong et al.(2016)]{kong16}Kong, X. L., Chen, Y., Guo, F., et al.\ 2016, \apj, 821, 32
\bibitem[Kong et al.(2012)]{kong12}Kong, X. L., Chen, Y., Li, G., et al.\ 2012, \apj, 750, 158
\bibitem[Kong et al.(2017)]{kong17}Kong, X. L., Guo, F., Giacalone, J., Li, H., Chen, Y. \ 2017, \apj, 851, 38
\bibitem[Kwon et al.(2014)]{kwon14}Kwon, R.-Y., Zhang, J., \& Olmedo, O. 2014, \apj, 794, 148
\bibitem[Kwon \& Vourlidas(2018)]{kwon18}Kwon, R.-Y., \& Vourlidas, A. 2018, JSWSC, 8, A08
\bibitem[Kumar et al.(2016)]{kumar16}Kumar, P., Innes, D. E., \& Cho, K. S. 2016, \apj, 828, 28
\bibitem[Lario et al.(2017)]{lario17}Lario, D., Kwon, R.-Y., Richardson, I. G., et al. 2017, ApJ, 838, 51
%\bibitem[Lee(1983)]{lee83}Lee, M. A. 1983, JGR, 88, 6109
%\bibitem[Lee(2005)]{lee05}Lee, M. A. 2005, \apjs, 158, 38
\bibitem[Li et al.(2018)]{li18}Li, X., Guo, F., Li, H., \& Li, S. 2018, ApJ, 866, 4
\bibitem[Li \& Lee(2015)]{li15}Li, G., \& Lee, M. A. 2015, ApJ, 810, 82
%\bibitem[Li et al.(2003)]{li03}Li, G., Zank, G. P., \& Rice, W. K. M.\ 2003, JGR, 108, 1082
\bibitem[Li et al.(2005)]{li05}Li, G., Hu, Q., \& Zank, G. P. 2005, in AIP Conf. Proc. 781, The Physics of Collisionless Shocks, ed. G. Li, G. P. Zank, \& C. T. Russell (Melville, NY:AIP), 233
\bibitem[Li et al.(2009)]{li09}Li, G., Zank, G. P., Verkhoglyadova, O. P., et al. 2009, \apj, 702, 998
\bibitem[Liu et al.(2019)]{liu19}Liu, Y. D., Zhu, B., Zhao, X. 2019, ApJ, 871, 8
\bibitem[Luhmann et al.(2018)]{luhmann18}Luhmann, J. G., Mays, M. L., Li, Y., et al. 2018, SpWea, 16, 557
\bibitem[Low(1986)]{low86}Low, B.~C.\ 1986, \apj, 310, 953
\bibitem[Ma et al.(2011)]{ma11}Ma, S., Raymond, J., Golub, L., et al.\ 2011, \apj, 738, 160
\bibitem[Mancuso \& Raymond(2004)]{mancuso04}Mancuso, S., \& Raymond, J. C.\ 2004, \aap, 413, 363
\bibitem[Mewaldt et al.(2005)]{mewaldt05}Mewaldt, R. A., Cohen, C. M. S., Labrador, A. W., et al. 2005, JGR, 110, A09S18
\bibitem[Mewaldt et al.(2015)]{mewaldt15}Mewaldt, R. A., Cohen, C., Mason, G., et al.\ 2015, Proc. 34th ICRC, 30
\bibitem[Mewaldt et al.(2012)]{mewaldt12}Mewaldt, R. A., Looper, M. D., Cohen, C. M. S., et al. 2012, SSRv, 171, 97
\bibitem[Nelson \& Melrose(1985)]{nelson85}Nelson, G. J., \& Melrose, D. B. 1985, in Solar Radiophysics, ed. D. J. McLean \& N. R. Labrum (Cambridge: Cambridge Univ. Press), 333
\bibitem[Parker(1965)]{parker65}Parker, E.~N.\ 1965, \planss, 13, 9
\bibitem[Pohjolainen et al.(2008)]{pohjolainen08}Pohjolainen, S., Pomoell, J., \& Vainio, R.\ 2008, \aap, 490, 357
\bibitem[Reames et al.(1997)]{reames97}Reames D. V., Kahler S. W., \& Ng C. K. 1997, ApJ, 491, 414
\bibitem[Reames(1999)]{reames99}Reames, D. V.\ 1999, \ssr, 90, 413
\bibitem[Reames(2009a)]{reames09a}Reames, D. V.\ 2009a, \apj, 693, 812
\bibitem[Reames(2009b)]{reames09b}Reames, D. V.\ 2009b, \apj, 706, 844
\bibitem[Reiner et al.(2003)]{reiner03}Reiner, M. J., Vourlidas, A., Cyr, O. C. St., et al.\ 2003, \apj, 590, 533
%\bibitem[Richardson et al.(2017)]{richardson17}Richardson, I. G., von Rosenvinge, T. T., Cane, H. V. 2017, AdSpR, 60, 755
\bibitem[Rouillard et al.(2016)]{rouillard16}Rouillard, A. P., Plotnikov, I., Pinto, R. F., et al. 2016, \apj, 833, 45
%\bibitem[Sandroos \& Vainio(2006)]{sandroos06}Sandroos, A., \& Vainio, R. 2006, \aap, 455, 685
%\bibitem[Sandroos \& Vainio(2009a)]{sandroos09a}Sandroos, A., \& Vainio, R.\ 2009a, \aap, 507, L21
%\bibitem[Sandroos \& Vainio(2009b)]{sandroos09b}Sandroos, A., \& Vainio, R.\ 2009b, \apjs, 181, 183
\bibitem[Schwadron et al.(2015)]{schwadron15}Schwadron, N. A., Lee, M. A., Gorby, M., et al.\ 2015, \apj, 810, 97
%\bibitem[Schwadron et al.(2015b)]{schwadron15b}Schwadron, N. A., Lee, M. A., Gorby, M., et al. 2015b, JPhCS, 642, 012025
\bibitem[Senanayake \& Florinski(2013)]{senanayake13}Senanayake, U. K., \& Florinski, V. 2013, \apj, 778, 122
\bibitem[Share et al.(2018)]{share18}Share, G. H., Murphy, R. J., White, S. M., et al. 2018, ApJ, 869, 182
\bibitem[Shen et al.(2013)]{shen13}Shen, C., Li, G., Kong, X., et al.\ 2013, \apj, 763, 114
\bibitem[Sokolov et al.(2004)]{Sokolov2004} Sokolov, I.~V., Roussev, I.~I., Gombosi, T.~I., et al.\ 2004, \apjl, 616, L171
\bibitem[Susino et al.(2015)]{susino15}Susino, R., Bemporad, A., Mancuso, S. 2015, ApJ, 812, 119
\bibitem[Tylka et al.(2005)]{tylka05}Tylka, A. J., Cohen, C. M. S., Dietrich, W. F., et al. 2005, \apj, 625, 474
\bibitem[Tylka \& Lee(2006)]{tylka06}Tylka, A. J., \& Lee, M. A. 2006, \apj, 646, 1319
\bibitem[Vainio et al.(2017)]{vainio17}Vainio, R., Raukunen, O., Tylka, A., et al. 2017, A\&A, 604, A47
\bibitem[Verkhoglyadova et al.(2015)]{verkhoglyadova15}Verkhoglyadova, O. P., Zank, G. P., \& Li, G.\ 2015, PhR, 557, 1
\bibitem[Wu \& Qin(2018)]{wu18}Wu, S. S., \& Qin, G. 2018, JGR, 123
\bibitem[Zank et al.(2000)]{zank00}Zank, G. P., Rice, W. K. M., \& Wu, C. C. 2000, JGR, 105, 25079
\bibitem[Zhao et al.(2016a)]{zhao16a}Zhao, L., Li, G., Mason, G. M., et al. 2016a, RAA, 16, 190
\bibitem[Zhao et al.(2016b)]{zhao16}Zhao, L., Zhang, M., \& Rassoul, H. K. 2016b, ApJ, 821, 62
\bibitem[Zhao et al.(2018a)]{zhao18a}Zhao, L. L., Adhikari, L., Zank, G. P., et al. 2018a, ApJ, 856, 94
\bibitem[Zhao et al.(2018b)]{zhao18b}Zhao, M. X., Le, G. M., \& Chi, Y. T. 2018b, RAA, 18, 74
\end{thebibliography}
\end{document}